\documentclass[aps,prd,twocolumn,nofootinbib,longbibliography,amsfonts,amssymb,amsmath,superscriptaddress]{revtex4-2}

\usepackage{array}
\usepackage[caption=false]{subfig}
\newcolumntype{L}{>{\centering}m{0.1\textwidth}}
\usepackage[english]{babel}
\usepackage{microtype}
\usepackage{comment}
\usepackage{graphicx}
\usepackage[normalem]{ulem}
\usepackage{tabularx}
\usepackage{tensor}
\usepackage[dvipsnames]{xcolor}
\usepackage{natbib}

\graphicspath{{./figs/}}

\usepackage[pdftex]{hyperref}

\begin{document}
	


\title{Nonlinear effects in the black hole ringdown: absorption-induced mode excitation}

\author{Laura Sberna}
\email{laura.sberna@aei.mpg.de}
\affiliation{Max Planck Institute for Gravitational Physics (Albert Einstein Institute) Am M\"{u}hlenberg 1, 14476 Potsdam, Germany}

\author{Pablo Bosch}
\affiliation{GRAPPA and Institute of High-Energy Physics, University of Amsterdam, Science Park 904, 1098 XH Amsterdam, The Netherlands}

\author{William E. East}
\affiliation{Perimeter Institute, 31 Caroline St N, Ontario, Canada}

\author{Stephen R. Green}
\affiliation{Max Planck Institute for Gravitational Physics (Albert Einstein Institute) Am M\"{u}hlenberg 1, 14476 Potsdam, Germany}

\author{Luis Lehner}
\affiliation{Perimeter Institute, 31 Caroline St N, Ontario, Canada}

\begin{abstract}
  Gravitational-wave observations of black hole ringdowns are commonly
  used to characterize binary merger remnants and to test general
  relativity. These analyses assume linear black hole perturbation
  theory, in particular that the ringdown can be described in terms of
  quasinormal modes even for times approaching the merger. Here we
  investigate a nonlinear effect during the ringdown, namely how a
  mode excited at early times can excite additional modes as it is
  absorbed by the black hole. This is a third-order secular effect:
  the change in the black-hole mass causes a shift in the mode
  spectrum, so that the original mode is projected onto the new
  ones. Using nonlinear simulations, we study the ringdown of a
  spherically-symmetric scalar field around an asymptotically anti-de
  Sitter black hole, and we find that this ``absorption-induced mode
  excitation'' (AIME) is the dominant nonlinear effect. We show that
  this effect takes place well within the nonadiabatic regime, so we
  can analytically estimate it using a sudden mass-change
  approximation. Adapting our estimation technique to
  asymptotically-flat Schwarzschild black holes, we expect AIME to
  play a role in the analysis and interpretation of current and future
  gravitational wave observations.

\end{abstract}

\maketitle

\section{Introduction}

Gravitational waves provide an unprecedented ability to witness the
approach to equilibrium of strongly-deformed black holes. Immediately
following a binary merger, the remnant is in a highly nonequilibrium
state, which emits gravitational radiation via characteristic
ringing. Observations of these waves by LIGO and Virgo can be used to
spectroscopically characterize the final black hole, to test
Einstein's theory of gravity in an extremely dynamical context, and to
establish consistency with no-hair
theorems~\cite{PhysRev.164.1776,Israel:1967za,PhysRevLett.26.331} and
the final state conjecture~\cite{1969NCimR...1..252P}.



To leading order in the deviation from a stationary black hole, the
ringdown is mostly\footnote{The ringdown can also contain an initial
  direct component, as well as a late-time power-law tail. We do not
  consider these components in this work.} characterized by a
collection of quasinormal modes (QNMs)~\cite{Nollert:1999ji,Kokkotas:1999bd,Berti:2009kk}. These have
complex frequency, corresponding to oscillations and decay. The QNM
spectrum is uniquely determined by the mass and spin of the black
hole, which makes ringdown analysis an especially exciting prospect
for gravitational-wave astronomy. Indeed, consistency between measured
quasinormal frequencies (QNFs) and predicted final states based on
inspiral measurements give rise to stringent tests of general
relativity~\cite{LIGOScientific:2016lio,LIGOScientific:2021sio}. For
particularly loud events, there is some observational evidence for
modes beyond the fundamental $(l,m)=(2,2)$, consistency of which
would test the no-hair theorem~\cite{Isi:2019aib,Capano:2021etf,cotestainprep}.

Although it is not yet possible to routinely observe multiple ringdown
modes for individual events, this will change with the enhanced
sensitivity of upgraded and future
detectors~\cite{Berti:2005ys,Maggiore:2019uih,Bailes:2021tot}. On the
theoretical side, high-accuracy signal modeling will be critical for
controlling systematic errors in parameter estimates and for
confidently constraining deviations from GR.  To that end, it will be
necessary to understand the ringdown beyond linear order---either to
incorporate nonlinear corrections into models or to show they are
negligible.

General relativity is a fundamentally nonlinear theory, and the
ringdown is no exception: QNMs can interact. This can be analyzed by
comparing linear theory against numerical relativity simulations. Some
examples include second-order effects in the ringdown following a
head-on black-hole collision~\cite{PhysRevLett.77.4483}, in the
black-hole response to strong incident
waves~\cite{Zlochower:2003yh,East:2013mfa}, mode-doubling during the
ringdown~\cite{Bantilan:2012vu,Ripley:2020xby}, and even
gravitational-wave turbulence in asymptotically anti-de Sitter (AdS)
spacetimes~\cite{Green:2013zba,Adams:2013vsa} (and possibly
asymptotically flat spacetimes~\cite{Yang:2014tla}). 

In contrast, for
typical mergers, sums consisting purely of (linear) QNMs have been
shown to provide a good fit to numerical relativity from
surprisingly early times\footnote{Models based on higher QNFs have
  been less successful in inferring remnant
  properties~\cite{Finch:2021iip,Li:2021wgz}.}~\cite{London:2014cma,Giesler:2019uxc,Cook:2020otn,
  Dhani:2021vac, Finch:2021iip,Li:2021wgz,Zertuche:2021xkb} (although
this does not automatically imply that the solution itself is
linear). It therefore remains to be understood to what degree nonlinear
mode interactions are present in astrophysical mergers: it has been
proposed that the formation of a common horizon ``hides''
nonlinearities from asymptotic observers~\cite{Okounkova:2020vwu}, or
that the black-hole potential ``filters'' nonlinear features out of
the signal. If true, this hints at a remarkable further property of
black holes: not only do they hide singularities, but also
nonlinearities!

Here we study nonlinear interactions among QNMs and uncover a key
effect that causes new modes to be excited. We begin with a toy-model
system consisting of an asymptotically AdS black hole and a perturbing
scalar field in spherical symmetry. This system allows for precise
numerical studies with initial data comprising a single QNM---enabling
us to track the excitation of new modes over time. We observe clear
QNM excitation during the ringdown that scales nonlinearly with the
initial data. The dominant effect that we uncover is as follows: (1)
as the seed mode decays, its absorption causes the black-hole mass to
grow; (2) this change to the background spacetime results in a
modified QNM spectrum; and finally (3) the remaining seed mode (no
longer a QNM of the final black hole) is projected onto the new
spectrum, resulting in excitation of additional modes. We call this
effect ``absorption-induced mode excitation'' (AIME). For all cases
studied, the QNM decay---and hence the change in the background
spacetime---is fast compared to the typical oscillation time
scale. AIME can therefore be estimated by taking the approximation of
a sudden (non-adiabatic) perturbation to the background, in analogy with quantum mechanics~\cite{Shankar:102017}.  We show that such analytic
estimates are in good agreement with simulations.

Guided by the intuition gained from the toy model, we estimate the
magnitude of AIME for gravitational perturbations of (asymptotically
flat) Schwarzschild black holes. We estimate the change in mass due to
absorption for a typical QNM excited in a merger, and then we project
this mode onto the spectrum of the late-time black hole. We find that
for realistic amplitudes, this gives percent-level corrections that
could be relevant for gravitational-wave observations. In addition to
AIME, QNMs can be excited via direct mode
coupling~\cite{Ripley:2020xby}, but the nature of our simulations
isolates the AIME effect. At the practical level, our work provides
guidance for future ringdown models and illuminates the transition to
the linear regime. It also clarifies the role of nonlinearities and
cautions against over-interpreting the success of fits that use a set
of linear modes.

Although conceptually clear, the calculations involved in carrying out
the mode projection are somewhat nontrivial. In particular, they have
required us to generalize (in the Schwarzschild case) the work of
Teukolsky and Press~\cite{TeuPress} to complex frequencies. (To our knowledge this is the
first time this has appeared.) We
also make use of a recently-presented finite bilinear form on
Teukolsky perturbations~\cite{Green:2019oaxaca,Stephen}: for QNMs this gives
orthogonality, but since the initial perturbation is not a QNM of
the final black hole, it can also be used to perform the
projection. We provide full details of these calculations.


This work is organized as follows. Sections
\ref{SADS:linear}--\ref{sec:NumericalSAdS} are devoted to the toy
model. In Sec.~\ref{SADS:linear} we review the theory of scalar QNMs
in Schwarzschild-AdS. In Sec.~\ref{SADS:nonlinear} we describe AIME,
and in Sec.~\ref{sec:NumericalSAdS} we describe our numerical studies
to elucidate the effect. In Sec.~\ref{sec:asymp_flat} we analyze AIME
for asymptotically-flat spacetimes, making use of the estimates we
validate earlier for Schwarzshild-AdS. We summarize our work and
discuss implications for ringdown analysis and modeling in
Sec.~\ref{sec:discussion}. We use units with $G=c=1$ throughout.

\section{Background: quasinormal modes in Schwarzschild-anti-de Sitter}\label{SADS:linear}
We study Einstein gravity with a negative cosmological constant coupled to a massless complex scalar field. The equations of motion are
\begin{align}
G_{ab} - \frac{3}{L^2} g_{ab} &= T_{ab}, \label{eq:AdSmotion1}\\
\nabla^a \nabla_a \phi &= 0 \label{eq:AdSmotion2},
\end{align}
where $ G_{ab} $ is the Einstein tensor, $L$ is the AdS length scale, and 
$ T_{ab} $ is the scalar field stress-energy tensor,
\begin{equation}\label{eq:scalarse}
T_{ab} = \frac{1}{16 \pi} \left[ \partial_{a} \phi^* \partial_{b} \phi + \partial_{a} \phi \partial_{b} \phi^*- g_{ab}  \partial^c \phi^* \partial_c \phi  \right] \, .
\end{equation}

We consider perturbations of the Schwarzschild-AdS (SAdS) metric,
\begin{equation}\label{eq:SAdS}
ds^2 = - f_{\rm SAdS}(r)dt^2 + f_{\rm SAdS}(r)^{-1} dr^2 + r^2 d\Omega_2^2,
\end{equation}
where
\begin{equation}
f_{\rm SAdS}(r) = 1 - \frac{2 M}{r} + \frac{r^2}{L^2}. 	
\end{equation}
We set $ L=1 $ throughout.

\subsection{Scalar quasinormal modes}
\label{sec:Scalar}

At leading order, the scalar field evolves in a fixed SAdS
background. The response to a small perturbation is characterized by
the scalar QNMs. With the ansatz
$ \phi = e^{-i\omega t} R(r) Y_l^m (\theta, \varphi) $, with $ Y_l^m $
a spherical harmonic, we obtain the radial equation,
\begin{equation}\label{key}
\frac{d}{dr} \left(r^2 f_{\rm SAdS}(r) \frac{d}{dr}R \right) + \left( \frac{\omega^2r^2}{f_{\rm SAdS}(r)} - l (l+1) \right) R = 0 \, .
\end{equation}
By redefining the radial function $ R(r)= r^{-1} X(r)$ and using the tortoise coordinate $ \frac{d r_*}{d r}  = f_{\rm SAdS}^{-1}(r)$, the equation is cast into a Schr\"{o}dinger-like form,
\begin{equation}\label{eq:Schrod}
\frac{d^2 }{d r_*^2}X + (\omega^2 - V_{\rm SAdS}(r)) X = 0,
\end{equation} 
where 
\begin{equation}\label{eq:SchrodV}
V_{\rm SAdS}(r) = - f_{\rm SAdS}(r) \left[\frac{l(l+1)}{r^2} + \frac{df_{\rm SAdS}(r)}{d r} \frac{1}{r} \right].	
\end{equation}
The potential $V_{\rm SAdS}$ asymptotes to zero at the horizon and
at spatial infinity. 
The horizon condition identifying scalar QNMs is the
standard ingoing condition,
\begin{equation}\label{eq:ingoingSAdS}
\phi \sim e^{-i \omega (t+r_*)}, \quad r_*\rightarrow-\infty .
\end{equation}
Near spatial infinity, the radial function has two solutions, $ \phi \sim {\rm constant} $ and $ \phi \sim r^{-3} $. We 
impose reflective boundary condition at the AdS boundary and therefore choose the solution $ \phi \sim r^{-3} $.
Last, we compute the scalar QNM frequencies and QNM radial profiles numerically using Leaver's method~\cite{Leaver,Gautschi} as detailed in Ref.~\cite{Bosch_2020}. Spherically-symmetric QNFs are simply labeled by the overtone number $ n $.

\subsection{Excitation coefficients}\label{sec:qnm_exc_linear}

The amplitudes and phases of QNMs are determined from the initial data
through the \emph{excitation
  coefficients}~\cite{Andersson:1996cm,Glampedakis:2003dn,Berti_2006}. These
coefficients naturally arise in the initial data problem for
perturbations of a black hole spacetime, in a theoretical
framework pioneered in Refs.~\cite{Leaver, PhysRevD.34.384,
  doi:10.1063/1.527130}. Here we adapt the formalism to the case of a scalar
perturbation of SAdS.

Re-introducing the scalar-field time dependence, we want to solve in full generality the equation 
\begin{equation}\label{key}
-\frac{d^2}{dt^2} X + \frac{d^2}{dr_*^2} X - V_{\rm SAdS}(r) X = 0,
\end{equation}
We define the Laplace transform of the field variable $ X $,
\begin{equation}\label{eq:transform}
\hat{X}(\omega, r) = \int_{0}^{+\infty} d t X(t,r) e^{+i\omega t} ,
\end{equation}
and the inverse transform 
\begin{equation}\label{eq:transform}
X(t, r) = \frac{1}{2 \pi} \int_{-\infty + i c}^{+\infty + i c} d \omega\hat{X}(\omega,r) e^{-i\omega t} , \quad c>0 .
\end{equation}
The transform satisfies
\begin{equation}\label{eq:Xhat}
\frac{d^2}{dr_*^2} \hat{X }(\omega,r)+ (\omega^2- V_{\rm SAdS}(r)) \hat{X}(\omega,r) = I(\omega,r),
\end{equation}
where the function $ I(\omega,r) $ is related to the value and derivative of the field at\footnote{The equations are invariant under time translation and we are therefore free to set the initial time to $ t=0 $.} $ t=0 $, $ \phi(0,r) = r^{-1}X^{(0)}(r) $ and $ \dot{\phi}(0,r) = r^{-1}\dot{X}^{(0)}(r) $. By the properties of the derivative of the transform,
\begin{equation}\label{key}
I(\omega,r) =  \left[ i \omega X^{(0)}(r) - \dot{X}^{(0)}(r) \right].
\end{equation}
The general solution to Eq.~\eqref{eq:Xhat} can be found via the Green's function of the homogeneous equation, defined in terms of two independent solutions $\hat{X}_{r_+}$, $\hat{X}_{\infty}$, 
\begin{align}\label{eq:sol1}
 \hat{X}_{r_+} \sim\begin{cases}
e^{-i\omega r_*}, \quad &{\rm as} \ r \rightarrow r_+, \\
B_{\infty}(\omega) r + \frac{A_{\infty}(\omega)}{r^2},  \quad  &{\rm as} \  r \rightarrow + \infty,
\end{cases}
\end{align}
and 
\begin{equation}\label{eq:sol2}
\hat{X}_{+ \infty} \sim \frac{1}{r^2}, \quad {\rm as} \ r \rightarrow + \infty \; .
\end{equation}
As we see below, it is not necessary to specify the behavior of $\hat{X}_{+ \infty}$ at the horizon.
We can then write the solution to Eq.~\eqref{eq:Xhat} as
\begin{align}\label{eq:Greens}
	\hat{X }(\omega,r) = &\; \hat{X }_{+ \infty}(\omega,r) \int_{-\infty }^{r_*} dr_*' \frac{\hat{X }_{r_+}(\omega,r_*') I(\omega,r_*') }{ W(\omega) } \nonumber \\
	&+  \hat{X }_{r_+}(\omega,r) \int_{r_*}^{0} dr_*' \frac{\hat{X }_{+ \infty}(\omega,r_*') I(\omega,r_*')}{ W(\omega) }.
\end{align}
The Wronskian of two independent solutions is independent of $r$, so we can choose to compute it at spatial infinity, $ W (\omega)= 3 B_{\infty}(\omega) $. 

\begin{figure}[t!]
	\begin{center}
		\includegraphics[trim={8.5cm 8.2cm 8.cm 9.cm},clip,width=0.98\linewidth]{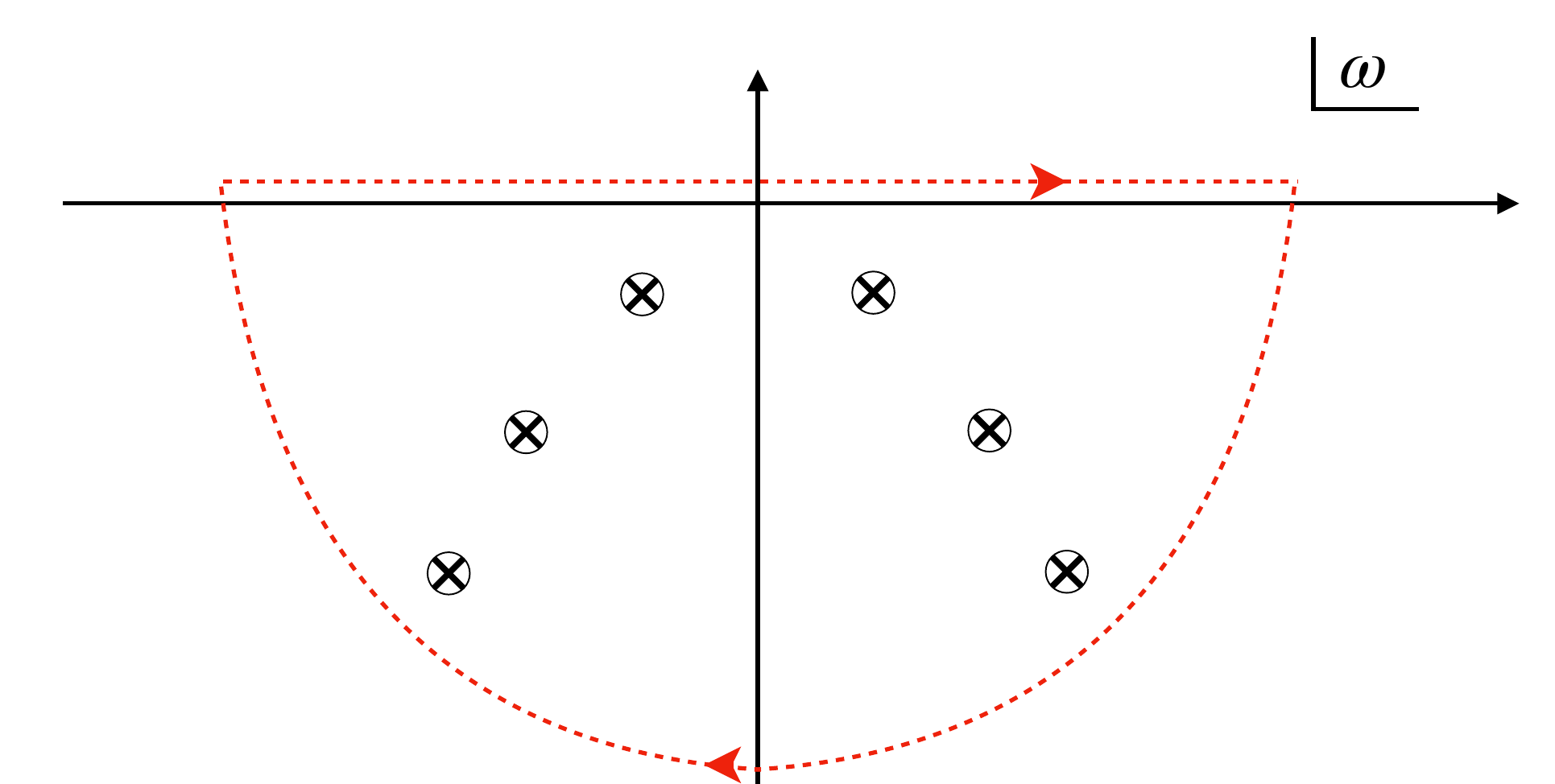}
	\end{center}
	\caption{Schematic integration contour used to evaluate the transform of \eqref{eq:Greens}. The integral is originally defined with a contour running above the real line from left to right. The crossed circles mark zeros of the Wronskian (i.e., the QN frequencies), while the semicircle at infinity carries a non-zero contribution associated with the direct propagation of the perturbation to the observer along null geodesics.
	} \label{fig:contour}
\end{figure}

To go back to the time domain, we need to plug this solution into the inverse transform \eqref{eq:transform} and integrate over $\omega$. 
The integrand has poles in the complex-$\omega$ plane at the zeros of the Wronskian, $ B_{\infty}(\omega)=0 $, as depicted in Fig.~\ref{fig:contour}. As seen by the asymptotic behavior in Eq.~\eqref{eq:sol1}, these are precisely the
QNFs $ \omega_n = \omega_{n,\rm R} + i \omega_{n,\rm I}$, with $ \omega_{\rm I}<0 $ and $n$ the overtone number.
 Thanks to the asymptotic behavior of the potential in the SAdS background, the integrand does not exhibit any branch cuts (present in asymptotically flat spacetimes).
 
 The inverse transform, whose integral runs along and above the real line, is equivalent to the integral along the closed contour drawn in Fig.~\ref{fig:contour} minus the integral along the semi-circle at infinity. 
In this work, we focus on the QNM contribution to the response (i.e., the poles), and neglect the contribution of the semicircle at large $ |\omega| $ (corresponding to the direct propagation of the perturbation to the observer). The latter is absent when the initial data extends to null infinity, as is the case in our simulations.
We are left with the integral on the closed contour, which amounts to the sum of the residuals associated with all the poles. Asymptotically, 
\begin{align}\label{key}
	X(t,r)  \sim \sum_n  C_n  \frac{1}{r^2} \, e^{- i \omega_n t} ,
\end{align}
at the AdS boundary, where the QNM \emph{excitation coefficients} are defined as
\begin{align}\label{eq:Cn}
	C_n = & \; i A_{\infty}(\omega_n) \left.\left(\frac{d W}{d \omega}\right)^{-1} \right|_{\omega_n} \nonumber \\
	& \times \left[\int_{-\infty}^{0} \mathrm{d}r_*' I (\omega_n,r_*') \hat{X}_{\omega_n}(\omega_n,r_*')  \right] .
\end{align}
and where we defined the QN radial profile $  \hat{X}_{\omega_n} =  \hat{X}_{\infty} ({\omega_n}, r_*)  =  \hat{X}_{r_+}({\omega_n}, r_*)/A_{\infty} $, so that $ r^2 \hat{X}_{\omega_n} = 1$ at the AdS boundary.  

The excitation coefficients are proportional to the overlap integral of the initial condition with the relevant QNM profile. The prefactors, which are independent of the initial condition and only depend on the properties of the background geometry and the type of perturbation considered, are known as the \emph{excitation factors},
\begin{equation}\label{eq:Bn}
B_n =  i A_{\infty}(\omega_n) \left.\left(\frac{d W}{d \omega}\right)^{-1} \right|_{\omega_n} .
\end{equation}

In this work, we are interested in how QNMs interact nonlinearly while being absorbed by the black hole. This is best explored using QNM initial data. For a QNM initial perturbation of frequency $\hat{\omega}$, the excitation coefficients read 
\begin{align}\label{eq:Cnomegas}
	C (\omega; \hat{\omega}) = & \; i A_{\infty}(\omega) \left.\left(\frac{d W}{d \omega}\right)^{-1} \right|_{\omega}  i( \omega + \hat{\omega}) \nonumber \\
	& \times \left[\int_{-\infty}^{0} \mathrm{d}r_*' \, \hat{X}_{\hat{\omega}}(\hat{\omega},r_*') \hat{X}_{\omega}(\omega,r_*')  \right] .
\end{align}

The overlap integral is absolutely convergent at the horizon only if the source has compact support. For a source extending to the horizon, as is the case for QNM initial data, a regularization procedure is needed to handle the divergence of the absolute value of the QN profile at the bifurcation surface. To this end, we implement a subtraction procedure (as discussed, e.g., in \cite{Green:2019oaxaca,Stephen} for Kerr),
\begin{align}\label{key}
	&\int_{-\infty}^{0} \mathrm{d}r_*' \, \hat{X}_{\hat{\omega}}(r_*') \hat{X}_{\omega}(r_*')\nonumber
	\\
	&  \rightarrow  \int_{R_*}^{0} \mathrm{d}r_*' \, \hat{X}_{\hat{\omega}}(r_*') \hat{X}_{\omega}(r_*')  + \frac{i }{\omega + \hat{\omega}}  \hat{X}_{\hat{\omega}}(R_*) \hat{X}_{\omega}(R_*)  
\end{align}
independent of the regularization point $ R_* \rightarrow -\infty $.

Thanks to the regularization, we verified that  QNMs with different overtone number are \emph{orthogonal}, $ C (\omega_n, \omega_{n'}) = 0 $ for $ \omega_n \neq \omega_{n'} $ (see again \cite{Green:2019oaxaca,Stephen} for Kerr). 
We show the convergence of the overlap integral and the orthogonality of a few modes in Appendix~\ref{app:convergence}.

\section{Absorption-induced mode excitations}\label{SADS:nonlinear}

Several studies have identified nonlinear behavior, in particular
turbulence, in asymptotically AdS black holes even under weak
perturbations~\cite{Carrasco:2012nf,Green:2013zba,
  Adams:2013vsa}. These studies demonstrated that, for sufficiently
large AdS black holes, there can be transfer of energy from short- to
long- wavelength metric perturbations (``inverse cascade''), which is
not captured by the standard QNM expansion. This phenomenology might
have an asymptotially-flat analog, for Kerr black holes near
extremality~\cite{Yang:2014tla}.

For generic black holes, the structure of the perturbative equations
suggests that first-order modes should drive, at second order, modes
with a characteristic combination of harmonics. This was identified
using numerical simulations in the gravity sector of asymptotically
flat~\cite{Zlochower:2003yh} and AdS black holes (in $5$
dimensions)~\cite{Bantilan:2012vu}. More recently, this second order
effect was observed with a perturbative numerical scheme in Kerr black
holes~\cite{Loutrel:2020wbw,Ripley:2020xby}.

At second order in perturbation theory, QNMs can also backreact on the
background spacetime, changing the black hole mass and spin. This
mediates a third order couplings between the modes. In this work, we
describe the absorption of QNMs by the black hole and the third order
mode coupling. In this section and the next, we focus on scalar QNMs
of AdS black holes. We consider gravitational perturbations of
asymptotically flat black holes in Sec.~\ref{sec:asymp_flat}.

\subsection{Absorption of quasinormal modes}\label{sec:theorybeyond}
Quasinormal modes have ingoing energy flux through the black hole horizon, which causes the black hole area to increase over time. 
Waves can be absorbed by the black hole horizon at all stages of a
binary merger. Incoming waves, however, need to overcome the potential
barrier outside the horizon. During the inspiral, most waves are
reflected by the barrier, and the effect of absorption is
suppressed~\cite{Poisson:1994yf,Bernuzzi:2012ku}. Indeed, absorption
appears at high orders in post-Newtonian (PN) expansions in the
orbital velocity $v_{\rm orb} \simeq (M f)^{1/3}$ ($M$ is the total
mass, $f$ the gravitational-wave frequency): 4PN in
Schwarzschild~\cite{Poisson:1994yf}, 2.5PN in
Kerr~\cite{Tagoshi:1997jy}. During the ringdown, however, $M f$ can be larger and the effect of absorption non-negligible. As we will see, black-hole absorption could be a
significant effect in the early ringdown, appearing at third order in
black hole perturbation theory.

To see this, we first compute the flux of QNMs across the black hole horizon. We write a generic spherically symmetric black hole metric in ingoing Eddington-Finkelstein coordinates,  
\begin{equation}\label{eq:metricEF}
ds^2 = - f dv^2 + 2 dv dr +r^2 d\Omega^2 ,
\end{equation}
where $v= t+r_*$.
Close to the horizon, a scalar QNM can be approximated as
\begin{equation}\label{eq:scalar_horizon}
\phi \sim A_{r_+} \, e^{-i \omega v} Y_l^m (\theta, \varphi) \, ,
\end{equation}
where $ A_{r_+} $ is the complex horizon amplitude. 
The time-like Killing vector
$t^a = 
 \partial_v^a - f \partial_r^a $ defines a conserved scalar field energy current,
\begin{equation}\label{key}
J^a = -T^{ab}t_b \, ,
\end{equation}
where the scalar-field stress-energy is given in Eq.~\eqref{eq:scalarse}. The integrated energy flux across the event horizon is 
\begin{equation}\label{key}
\frac{d M}{d v} = \Phi (v)= \int_{\rm horizon} d\Omega \, n_a J^a
\end{equation}
where $ n_a = -t_a $ is the inward normal to the horizon,\footnote{We follow Wald's convention~\cite{wald1984general}. 
} and $M$ is the black hole mass (technically, the Misner-Sharp mass at the horizon~\cite{MisnerSharp,Cahill1970}). 
Using the asymptotic limit of the scalar field~\eqref{eq:scalar_horizon}, one finds that for a single mode~\cite{Dolan:2007mj},
\begin{equation}\label{eq:fluxgeneric}
\Phi (v) = \frac{   |\omega|^2 }{8 \pi} |A_{r_+}|^2 e^{2 v \omega _{\rm I}},
\end{equation}
for any $(l,m)$. For a perturbation containing modes up to overtone $N$, this generalizes to
\begin{equation}\label{eq:fluxgeneric_N}
\Phi (v) =  \frac{  1 }{8 \pi} \sum_{n_i,n_j=0}^N (A_{r_+,n_i} A^*_{r_+,n_j}) (\omega _{n_i} \omega^*_{n_j}) e^{- i v \omega _{n_i}} e^{+ i v \omega^*_{n_j}}.
\end{equation}
Perturbations with different $(l,m)$ do not mix at this order in the flux formula, thanks to the orthogonality of spherical harmonics.\footnote{This statement extends to higher-spin perturbations of Schwarzschild and SAdS. The spin-weighted spheroidal harmonics in Kerr are not orthogonal, so different $(l,m)$ modes will mix in the flux. We will come back to this in Section~\ref{sec:asymp_flat}.}

Equation~\eqref{eq:fluxgeneric} confirms that QNMs have a positive energy flux through the horizon. 
Energy falls into the black hole at an exponential (decaying) rate, with a time scale given by half of the mode decay time, 
\begin{equation}
	 \tau_{\rm back.} \simeq (2 \omega_{\rm I})^{-1} .
\end{equation}
This result is valid for scalar QNMs of any spherically-symmetric static black hole. The analogous calculation of the flux of angular momentum can be found in Appendix~\ref{app:dJdM}.
The energy flux is also proportional to the square of the scalar field amplitude at the horizon, which is itself proportional to the amplitude observed far away from the black hole, although the proportionality constant can be nontrivial. 

To see how the black hole horizon grows over time as a result of the QNM flux, we write the black hole area $ \mathcal{A} $ in terms of the irreducible mass,
\begin{equation}\label{key}
\mathcal{A}= 16 \pi M_{\rm irr}^2 \, .
\end{equation}
%
In SAdS, the black-hole mass and the irreducible mass are related by $ M = M_{\rm irr} (1+4 M_{\rm irr}^2 L^{-2} )  $. 
The area evolution is then given by
\begin{equation}\label{key}
\frac{d\mathcal{A}}{dv} = 32 \pi M_{\rm irr} \frac{d M_{\rm irr}}{d M} \Phi  \quad {\rm in \ SAdS,} 
\end{equation}
where $d M_{\rm irr}/d M>0$. This implies
\begin{equation}\label{eq:deltaASAdS}
\mathcal{A}(v) = \mathcal{A}(0) + \frac{2 M_{\rm irr} |\omega |^2  \left| A_{r_+}\right| ^2}{\omega_{\rm I}} \frac{d M_{\rm irr}}{d M} \left(e^{2 v \omega _{\rm I}} -1\right)\, ,
\end{equation}
where $\mathcal{A}(\infty)>\mathcal{A}(0)$. The total area change from $v=0$ is given by
\begin{equation}\label{eq:delta_area}
\Delta \mathcal{A} = - \frac{2 M_{\rm irr} |\omega |^2  \left| A_{r_+}\right| ^2}{\omega_{\rm I}} \frac{d M_{\rm irr}}{d M} \; .
\end{equation}
The expressions above can also be written in terms of $t$, the time measured by an observer on the same null slice at the AdS boundary (see Fig.~\ref{fig:penrose_SAdS}).

\subsection{Heuristics of mode excitation via absorption}\label{sec:QNMs_nonlinear} 

The evolution of the black-hole background can affect the ringdown in
two ways---by altering the QNM \emph{spectrum} and the mode
\emph{content} of the ringdown (the excitation of modes). We can
understand these effects in two extremes:
\begin{enumerate}
	\item the \emph{adiabatic regime}, when the background evolves very slowly compared to the typical dynamical time of QNMs already excited, $\tau_{\rm back.} \gg \tau_{\rm osc.}$. We associated the dynamical timescale of a QNM with its real frequency, i.e., $\tau_{\rm osc.} = 2 \pi / \omega_{\rm R}$;
	\item the \emph{nonadiabatic regime} (also known as ``diabatic"), when the background evolves rapidly compared to the modes, $\tau_{\rm back.} \ll \tau_{\rm osc.}$.
\end{enumerate}
This distinction was previously explored in AdS black holes in Ref.~\cite{Buchel:2013lla}. The authors found universal behavior in the black hole response to arbitrarily fast perturbing quenches, i.e., impulsive interactions with a scattering packet of energy, as well as adiabatic behavior in the opposite regime.

In the \emph{adiabatic regime}, the perturbation remains in the same QN state, and the mode content does not change. This is analogous to adiabatic processes in quantum mechanics~\cite{Shankar:102017}. If the initial data is composed of a single seed QNM, the mode evolves into the corresponding mode in the spectrum of the final black hole (with the same angular and overtone numbers). In the end, no other modes of the late-time spectrum are excited.

In the \emph{nonadiabatic regime}, modes feel a near-instantaneous
shift in the background. In the new background, the seed mode with
$n=\hat{n}$ is no longer a pure QNM, and therefore consists of a
collection of modes of the new black hole. This results in the
excitation of additional modes of the new spectrum (with
$n\neq \hat{n}$), a process we call ``absorption-induced mode
excitation", or AIME.

To re-expand the old mode in terms of the new ones, we compute
the excitation coefficients~\eqref{eq:Cnomegas} of the new modes with $\omega  = \omega_n(M_{\rm final})$, with initial data consisting of the original mode $\hat{n}$ with $\hat{\omega} =\omega_{\hat{n}} (M_{\rm init})$, 
\begin{equation}
	C^{\rm nonadiab.}_{n} = C(\omega_{\hat{n}}(M_{\rm init}),\omega_n(M_{\rm final})) \; .
\end{equation}
The excitation coefficients should approximate the AIME amplitudes. This re-expansion sources all overtones $n$ of the new spectrum, as QNMs of different black holes are not orthogonal. 

In a realistic scenario, we will have something in between: modes will be excited with a fraction of the near-instantaneous excitation coefficient~\eqref{eq:Cnomegas}, and their frequencies will adjust to the background evolution. As we will see in Section~\ref{sec:linear_sim}, the near-instantaneous excitation coefficients are a good approximation even when the system is not in the fully nonadiabatic regime.  

AIME is obviously nonlinear in nature: the amplitude of the excitations is cubic in the initial perturbation. We can understand this as follows. In the fully nonadiabatic regime, we need to project the initial perturbation (the $\hat{n}$ mode of a SAdS spacetime with $ M_{\rm irr}=M_{\rm irr,init} $) onto all the modes of the final spacetime, having mass $ M_{\rm irr}=M_{\rm irr, final} $. From Eq.~\eqref{eq:deltaASAdS}, we expect that at leading order the change in mass will be proportional to the square of the mode amplitude at infinity: $ \Delta M_{\rm irr} = M_{\rm irr, final} - M_{\rm irr, init} \sim |A_{\infty, \hat{n} }|^2 $. The amplitude of each excited mode $n$ will be given by its excitation coefficient~\eqref{eq:Cnomegas}. If the difference between the initial and final mass is equal to zero, the excitation coefficient of any mode $n\neq \hat{n}$ vanishes due to their orthogonality. We therefore expect the excitation coefficient corresponding to $n\neq \hat{n}$ to be proportional to $\Delta M$, to leading order in the perturbation amplitude. Taking into account that the excitation coefficient is also proportional to the amplitude of the initial data, we find
\begin{equation}\label{eq:thecubic}
A_{ \infty} \sim A_{ \infty, \hat{n} } \, \Delta M \sim A_{ \infty, \hat{n} }^3 .
\end{equation}
for $\hat{n}\neq n$. 
We will confirm this expectation with fully nonlinear calculations in the next section.

\section{Numerical results} \label{sec:NumericalSAdS}
In this section, we numerically
evolve a scalar field coupled to gravity 
to investigate nonlinear effects in the ringdown of a spherically symmetric
black hole in asymptotically AdS with reflecting boundary conditions.
We do so in two steps:
\begin{enumerate}
  \item In Sec.~\ref{sec:nonlinear_sim}, we numerically evolve the full system in spherical symmetry. We choose initial data containing a single scalar QNM with
  $\hat{n}>0$, which is ideally suited to study nonlinear behavior. This is because,
  after some time, nonlinearly excited modes with $n<\hat{n}$ will come to
  dominate the ringdown, simply because of their longer decay times (despite their initially small amplitudes). We observe nonlinearities in all of our simulations, for a variety of perturbing overtone
  numbers and amplitudes. These simulations confirm that the horizon area grows as a result of a QNM flux as predicted by Eq.~\eqref{eq:deltaASAdS}, and suggest
  that the dominant nonlinear process is AIME, as defined in Sec.~\ref{sec:QNMs_nonlinear}. 

  \item In Sec.~\ref{sec:linear_sim}, we further show that these nonlinear
excitations are independent of how we generalize QNM initial data to a
nonlinear setting. Moreover, we pin down the mechanism behind the nonlinear
excitations. We do so by using linear simulations of a scalar field in an evolving background designed to mimic the backreacted black hole spacetime. 
We show that these simulations, which single out the effect of absorption, are
an excellent approximation to the fully nonlinear runs. Linear simulations also
allow us to explore AIME in the adiabatic limit.
\end{enumerate}

\subsection{Numerical and fitting setup}
\label{sec:numerical_fitting_setup}

\begin{figure}[t]
	\begin{center}
		\includegraphics[trim={11.cm 12.cm 10.5cm 0.cm},clip,width=0.85\linewidth]{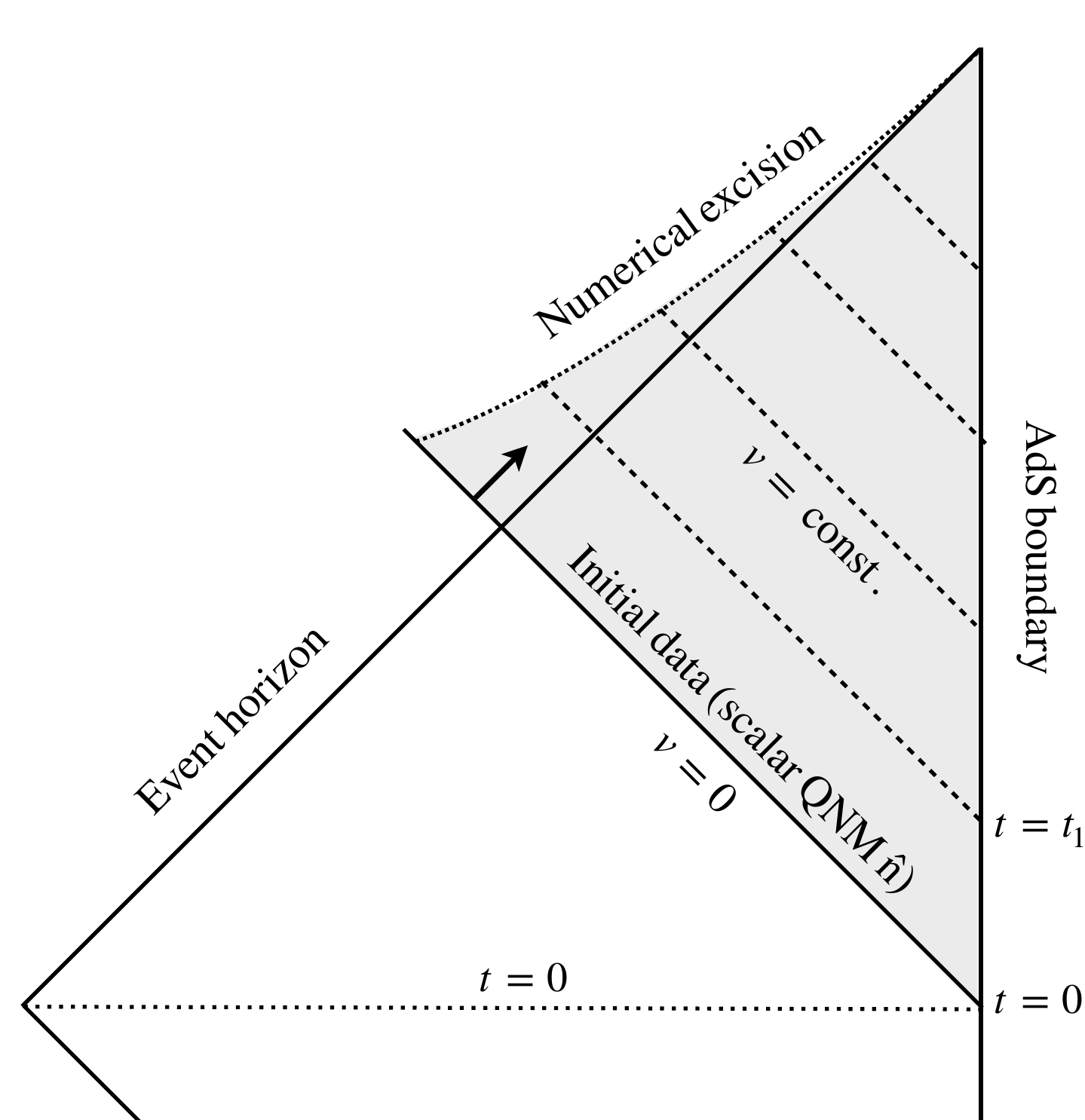}
	\end{center}
	\caption{ Schematic Penrose diagram showing the causal structure of Schwarzschild-AdS and the coordinates used. The domain of evolution of our numerical code is limited to the shaded area. 
	} \label{fig:penrose_SAdS}
\end{figure}

\subsubsection{Numerical Method}
We solve the full system of
equations~\eqref{eq:AdSmotion1}--\eqref{eq:AdSmotion2} in spherical symmetry
using the numerical code developed
in Ref.~\cite{Bosch_2016,Bosch_2020}, which is based on a
characteristic approach~\cite{Chesler:2013lia}. We restrict
to spherical symmetry, using the following ansatz for the metric
in ingoing Eddington-Finkelstein coordinates, 
\begin{equation}\label{key}
ds^2 = - A(v,r) dv^2+2 dv dr+\Sigma(v,r) d\Omega_2^2.
\end{equation}
The code uses a finite difference scheme with mixed second and fourth order
radial derivative operators satisfying summation by parts
\cite{Calabrese:2003yd,Calabrese:2003vx}, and fourth order Runge-Kutta for
the time evolution. 

We compactify the spatial domain by using the inverse radial
coordinate $\rho = 1/r$. This allows us to include the AdS boundary
$\rho=0$, so the computational domain is $0 \le \rho \le 1/r_0 $, with
the inner radius $r_0$ lying several grid points within the apparent
horizon, and thereby excising the singularity
(Fig.~\ref{fig:penrose_SAdS}). Equations are discretized
using a uniform grid throughout the compactified spatial domain. For
the results below, we typically use 4801 points to cover the domain in
$\rho$, and set $dv/d\rho=0.4$ for numerical stability. (See
Appendix~\ref{app:fits} for truncation error estimates.)
Boundary data consist of the ADM mass $M=0.104$, and the initial data is fully
specified by the scalar field profile on the initial null slice
$\phi(v=0)$.

The system is solved by integrating radially, starting from the AdS boundary,
along $v=\mathrm{constant}$ null curves to obtain the remaining field values
and their derivatives; the scalar field is then integrated one step
$\mathrm{d}v$ forward in time, and the procedure is iterated. 

%

In this work, we choose a specific QNM with overtone number $\hat{n}$ as initial data for
the scalar field. The radial QNM profile is obtained by specifying the
target overtone frequency and solving the radial Klein-Gordon equation on SAdS, following the methods
detailed in Ref.~\cite{Bosch_2020}. The radial profiles $\phi_{\hat{n}}(r)$ are
normalized using the maximum norm
\begin{equation}
\phi_{\mathrm{init. \; data}}(r) = A_{ \hat{n}} \frac{\phi_{\hat{n}}(r)}{\underset{r}{{\rm max}}\, \phi_{\hat{n}}(r) }\;,
\end{equation}
with $A_{ \hat{n}}$ setting the overall amplitude of the field.

For large amplitudes $A_{\hat{n}}$, a significant fraction of the ADM
mass $M$ will be contained within the scalar field itself. As a
consequence, the black hole will have a mass significantly 
smaller than $M$ and the actual black hole background will not precisely match the SAdS background for which the initial data were constructed. With this in mind, we can construct two types of initial data: 
\begin{itemize}
\item For ``global-mass'' initial data, we obtain the radial scalar
  field profiles for the target overtone $\hat{n}$ on a SAdS
  background of mass $M$. Then, this is used as initial data for the
  scalar field without updating the ADM mass $M$ of the
  spacetime. Thus, the perturbation frequency and profile matches the
  final black hole mass (after the perturbation has been completely
  absorbed) but has a small discrepancy with the black hole mass at
  the beginning of the evolution.
\item For ``matched'' initial data, we follow the same procedure
  outlined above and extract the irreducible mass of the black hole on
  the first slice $v=0$. Next, we use this extracted mass to compute
  the \emph{new} background mass, update the target overtone frequency
  $\omega_{\hat{n}}$, and finally obtain the \emph{updated} radial
  profiles. This process can be iterated, however we observe that with
  a single iteration we achieve very good agreement between the mass
  used for the initial data calculations and the mass computed on the
  first slice of the evolution.  These initial data should therefore
  better represent a single QNM of the initial black hole.
\end{itemize}

The asymptotic expansion of the scalar field about infinity, with reflecting
boundary conditions, takes the form~\cite{Bosch_2016,Bosch_2020}
\begin{equation}
\phi(v,r) = \frac{\phi_\infty(v)}{r^3} + O\left( \frac{1}{r^4}\right) .
\end{equation}
The quantity $\phi_{\infty}(v)$ is an output of the simulation, and contains
the information about the mode content of the solution. We extract it at the
AdS boundary as a time series $ \phi_{\infty}(v)$. Moreover, the time
coordinates $v$ and $t$ coincide at the AdS boundary; hence, unless otherwise
stated, we will refer to this time series $\phi_{\infty}(v)$ as $\phi(t)$ for
simplicity.  Another useful output of the simulation that we extract is the apparent
horizon area time series $\mathcal{A}_{\rm H}(v)$, which we denote
$\mathcal{A}(v)$. 



\subsubsection{Fitting Method}
We analyze the QNM content of the scalar field ringdown by fitting the time series with the model
\begin{align}\label{eq:fitmodelAdS}
\phi^{N_+,N_-}(t) =& \sum_{n=0}^{N_+-1} A_{+n} e^{-i \omega_n (t-t_0) + \varphi_{+n}} \nonumber \\&+ \sum_{n=0}^{N_--1} A_{-n} e^{-i \omega_{-n} (t-t_0) + \varphi_{-n}}.
\end{align}
containing $N_+$ and $N_-$ positive and negative frequency modes, respectively. The latter are sometimes referred to as mirror modes~\cite{Berti:2005ys}, and satisfy $ \omega_{-n} = - \omega_{n}^* = - \omega_{n,{\rm R}} + i \omega_{n,{\rm I}}  $, with $ \omega_{n,R}>0 $.

We determine the value of the amplitudes $ A_{\pm n} $ and phases $ \varphi_{\pm n} $ using a least squares fit to the boundary data, while holding the frequencies fixed. The QNFs used for the fit are computed based on the \emph{late-time} mass of the black hole, using Leaver's method with the code presented in Ref.~\cite{Bosch_2020} using the. The fits are performed between an early time $ t_0 $ and a late time $ T $. We assess the goodness of the fits by plotting the fit residuals and computing the mismatch between the model and the data, as described in Appendix~\ref{app:fits}.

\subsection{Nonlinear evolution}\label{sec:nonlinear_sim}

\begin{figure}[t]
	\centering
		\includegraphics[trim={0.9cm 1.cm .5cm 0.9cm},clip,width=0.98\linewidth]{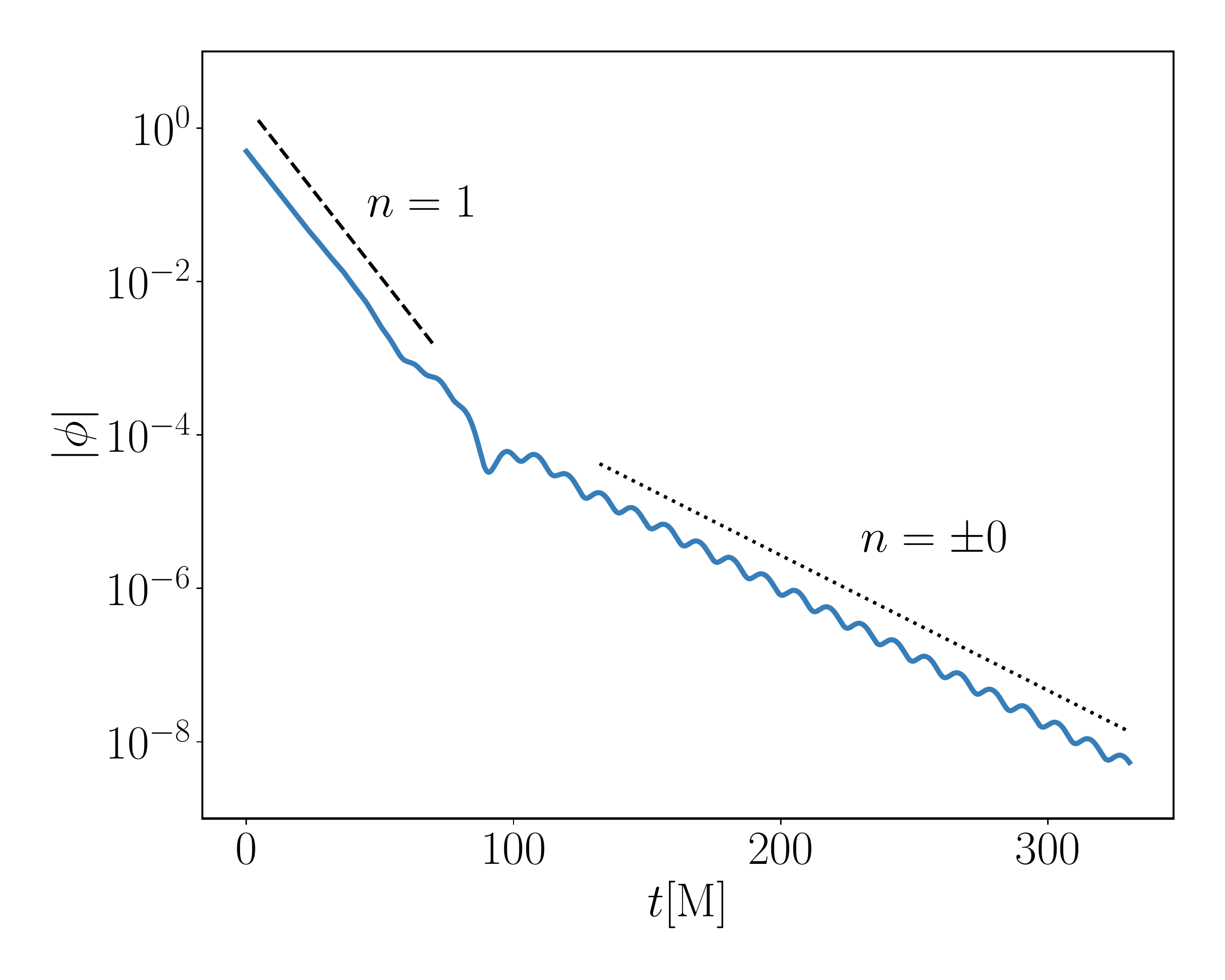}
	\caption{ Nonlinear scalar field ringdown for a QNM perturbation of overtone number $ \hat{n}=1 $ and amplitude $ A_{\hat{n}}=0.1 $. The time evolution of the absolute value of the scalar field reveals the presence at late times of two additional modes ($ n=\pm0 $).  
	For visual comparison, we plot the predicted slopes of the fundamental mode (dotted) and first overtone (dashed) associated with the ADM mass M.
	} \label{fig:data_resolution}
\end{figure}

The typical nonlinear response of the scalar field to QNM initial data
is exemplified by the ringdown shown in
Fig.~\ref{fig:data_resolution}. The response begins with the
perturbing mode ringing (in this case the first overtone,
$\hat{n}=1$), with no sign of an ``immediate response" phase. This can
be explained by the fact that the initial data is a QNM defined on the
entire ingoing null slice, from the AdS boundary up to the black hole
horizon, as opposed to compactly supported initial data.  At
intermediate times, we observe a transition where eventually the decay
of the first overtone causes it to drop below the fundamental, which
takes over for the rest of the evolution. At late times, as expected
for asymptotically AdS black holes, we do not detect a power-law
tail. We therefore use a pure QNM model \eqref{eq:fitmodelAdS} to fit
the data from $t_0 = 0$ and $T=300 M$.


Eventually, after the transition, the dynamics are dominated
by a mix of the positive and negative frequency fundamental modes, as seen
in the oscillating decay from $t=110M$ onwards in
Fig.~\ref{fig:data_resolution}.  These modes are longer-lived and smaller
in amplitude compared to the perturbing first overtone ($\hat{n}=1$).  This
simple three-mode model suggested by visual inspection is confirmed by a
least squares fit; see Fig.~\ref{fig:residuals} and Table~\ref{tb:table1}
in Appendix~\ref{app:fits}. The residual with just the $ n=1$ included in
the fit is well above the numerical error, but significantly decreases
with the inclusion of the $ n=+0 $ and $ n=-0 $ modes in the model. 
The asymmetric excitation of the positive and negative
fundamental modes should not come as a surprise; although the spectrum and
the radial profiles are symmetric under $ \omega \rightarrow -\omega^*$, $
X\rightarrow X^* $ (see Eqs.~\eqref{eq:Schrod}, \eqref{eq:SchrodV}), so
that $ X_{\omega} = X^*_{-\omega^*} $, our choice of positive frequency
initial conditions picks out a preferred sign.

The evidence for overtones above the perturbing mode, $n>1 $, or
including $n=-1$, is very weak. The residuals do not improve
significantly when adding these modes. In particular, residuals never
improve significantly at early times, perhaps due to the fact that our
fit model does not allow for evolution of the quasinormal frequencies
and amplitudes.  Furthermore, the amplitudes of the first three modes
remain consistent with the minimal model as overtones are added, but
the amplitudes of the higher tones are unstable; for example, the
$ n=2 $ mode amplitude doubles when adding the third
overtone. Similarly, the mismatch decreases significantly when
reaching the minimal model, but fluctuates around the minimum value as
overtones are added. Generically, nonlinear theory does not prohibit
the excitation of higher overtones, hence they are likely present in
the nonlinear data at late times. These higher overtones are hard to
identify due to a combination of perturbatively small amplitudes and
short decay times. For this reason, we limit the scope to the
minimal fit models that include $ n < n_{\mathrm{pert}} $ overtones in
the remainder of the section. We justify this choice further in
Appendix~\ref{app:fits}.

We verify that modes are excited nonlinearly by studying the excitations as a function of the QNM perturbation amplitude. We find that mode amplitudes behave cubically, as predicted in Eq.~\eqref{eq:thecubic},
\begin{equation}\label{key}
A_{n\neq\hat{n}} \sim A_{\hat{n}}^3,
\end{equation}
with $O(0.1-10) $ prefactors. We show this in
Fig.~\ref{fig:amplitudes} for $ \hat{n}=1 $ perturbations and in
Appendix~\ref{app:n2} for $ \hat{n}=2 $ perturbations, but we confirmed
this pattern\footnote{At higher $ \hat{n} $, the number of parameters to
fit increases as $ 4 (\hat{n}-1) $, and we are less confident in the
amplitudes of high-$ \hat{n} $ overtones recovered by the fit.} up to $
\hat{n}=10 $. We verified that modes are excited cubically also in larger
black holes, with $ r_+ \gg L $. We omit these results for brevity, and
because for larger black holes the late time data is more susceptible to
numerical error. 

\begin{figure}[t]
	\centering
		\includegraphics[trim={0.95cm 1.3cm .7cm 0.9cm},clip,width=0.98\linewidth]{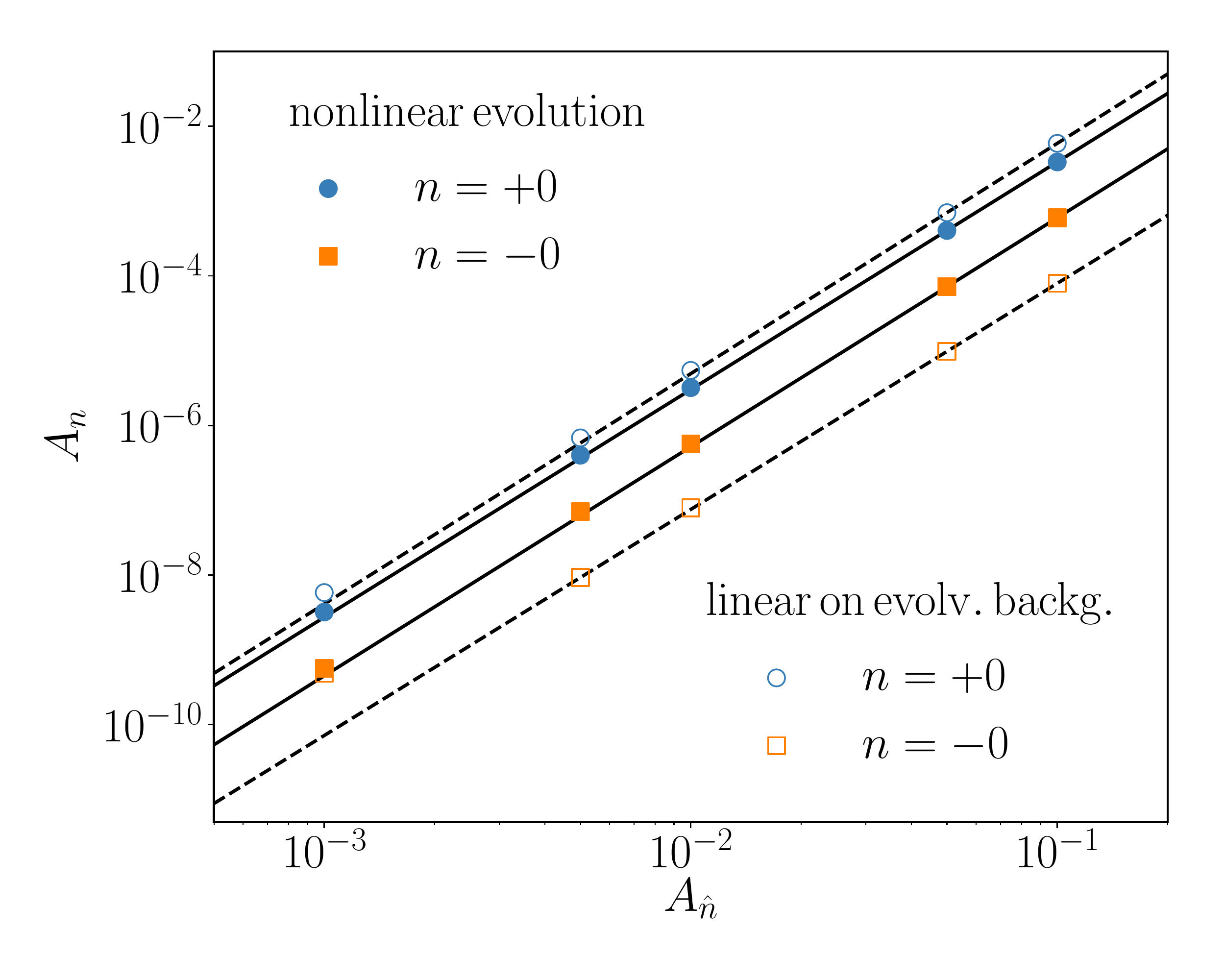} 
	\caption{ Amplitude of excited fundamental modes in nonlinear simulations (full circles) as a function of the $ \hat{n}=1 $ perturbation amplitude, and a power law fit (black, solid). We find a power of $3.04$ and $3.06$ for the $+0$ mode and the $-0$ mode, respectively. We also show the excitation amplitudes in the \emph{linear} simulations with evolving background (empty circles), and their power law fit (black, dashed), returning $3.08$ and $3.02$ for the power, respectively (when excluding the point with lowest amplitude). Excitations are cubic in the perturbation amplitude, in both nonlinear and linear simulations on a time-dependent background. The amplitudes deviate from the cubic relation when the perturbation becomes so small that excitations are comparable to the numerical error. 
	} \label{fig:amplitudes}
\end{figure}

\begin{figure}[t]
	\centering
	\includegraphics[trim={0.8cm 1.2cm 0.2cm 1.cm},clip,width=0.98\linewidth]{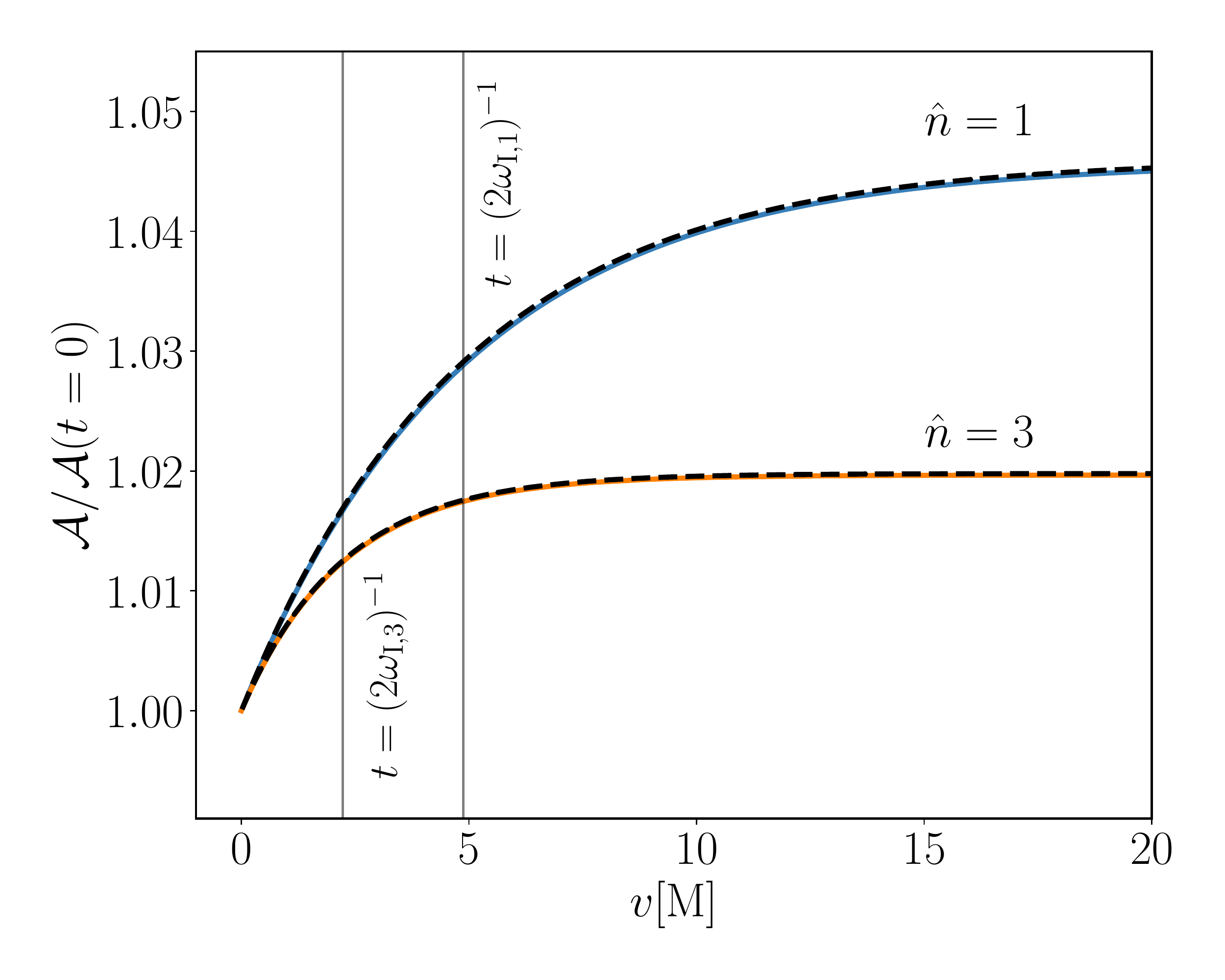}
	\caption{Area increase as a function of the Eddington-Finkelstein coordinate on the horizon. We show the area evolution in two simulations, with first and third overtone perturbations of amplitude $ A_{\hat{n}}=0.05 $. The analytic prediction is overlaid as a dashed line. 
	} \label{fig:area_pred}
\end{figure}

As discussed in Sec.~\ref{sec:theorybeyond}, we expect QNMs to backreact on the black hole, increasing its horizon area over time. Assuming that the perturbing mode is the dominant source of flux across the horizon, we can use Eq.~\eqref{eq:deltaASAdS} to predict the change in area. We simply read off the perturbing mode horizon amplitude (appearing in Eq.~\eqref{eq:deltaASAdS}) from its numerically determined profile, as well as the black hole area on the initial slice. The result, shown in Fig.~\ref{fig:area_pred} for $ \hat{n}=1 $ and $ \hat{n}=3 $ perturbations, is in excellent agreement with the numerical evolution of the horizon area, and shows an area variation of $ 2 $--$ 4 \%$. 

The agreement between the numerical and analytic area evolution confirms that its timescale is set by half the decay time of the dominant QNM in the ringdown. We can then use this timescale to determine whether the system will behave adiabatically or not. The comparison between the area evolution timescales and the mode oscillation frequencies (all shown in Table~\ref{tb:table_AdStimes}) tells us that all our simulations fall under the nonadiabatic regime, as defined in Sec.~\ref{sec:QNMs_nonlinear}. 

\begin{table}[t]
	\begin{ruledtabular}
		\begin{tabular}{ccc}
			n & $\tau_{\rm back.} [M]$ &$\tau_{\rm osc.} [M]$ \\ 
			\colrule
		$0$    &     12.33 & 24.41  \\ 
$1$     &    4.87 & 14.84  \\ 
$2$     &     3.05 & 10.55     \\ 
$3$     &     2.23 & 8.16  \\ 
		\end{tabular}
	\end{ruledtabular}
\caption{Timescale of backreaction $\tau_{\rm back.}=1 / 2\omega_I$ and oscillation $\tau_{\rm osc.}=2 \pi /\omega_R$ of the first four scalar modes in SAdS. Times are given in units of the BH mass, fixed to M=0.104. The backreaction timescale of our prototypical $n=1$ mode is shorter than the oscillation timescales of the modes up to at least $n=3$. This is also true for the other modes used as initial data in this work, $n=2$, $3$. Based on this estimate, their nonlinear excitations should behave non-adiabatically. }\label{tb:table_AdStimes}
\end{table}


\subsection{Linear evolution with time-dependent background}\label{sec:linear_sim}

The nonlinear simulations contain the full dynamics and mode
excitations produced by the scalar field backreaction, with the
expected cubic relation. To further isolate this effect we carry out a
special linear simulation, where we evolve the scalar field on a
time-dependent background. The spacetime is prescribed analytically to
have a evolving mass corresponding to the dominant QNM flux through
the horizon, as in Eq.~\eqref{eq:deltaASAdS}.  These linear
simulations single out the effect of the absorption-induced mode
excitation, allowing us to verify where the cubic excitation is
actually stemming from.

\begin{figure}[t]
	\centering
	\includegraphics[trim={1.2cm 1.cm .9cm 0.9cm},clip,width=0.9\linewidth, draft=False]{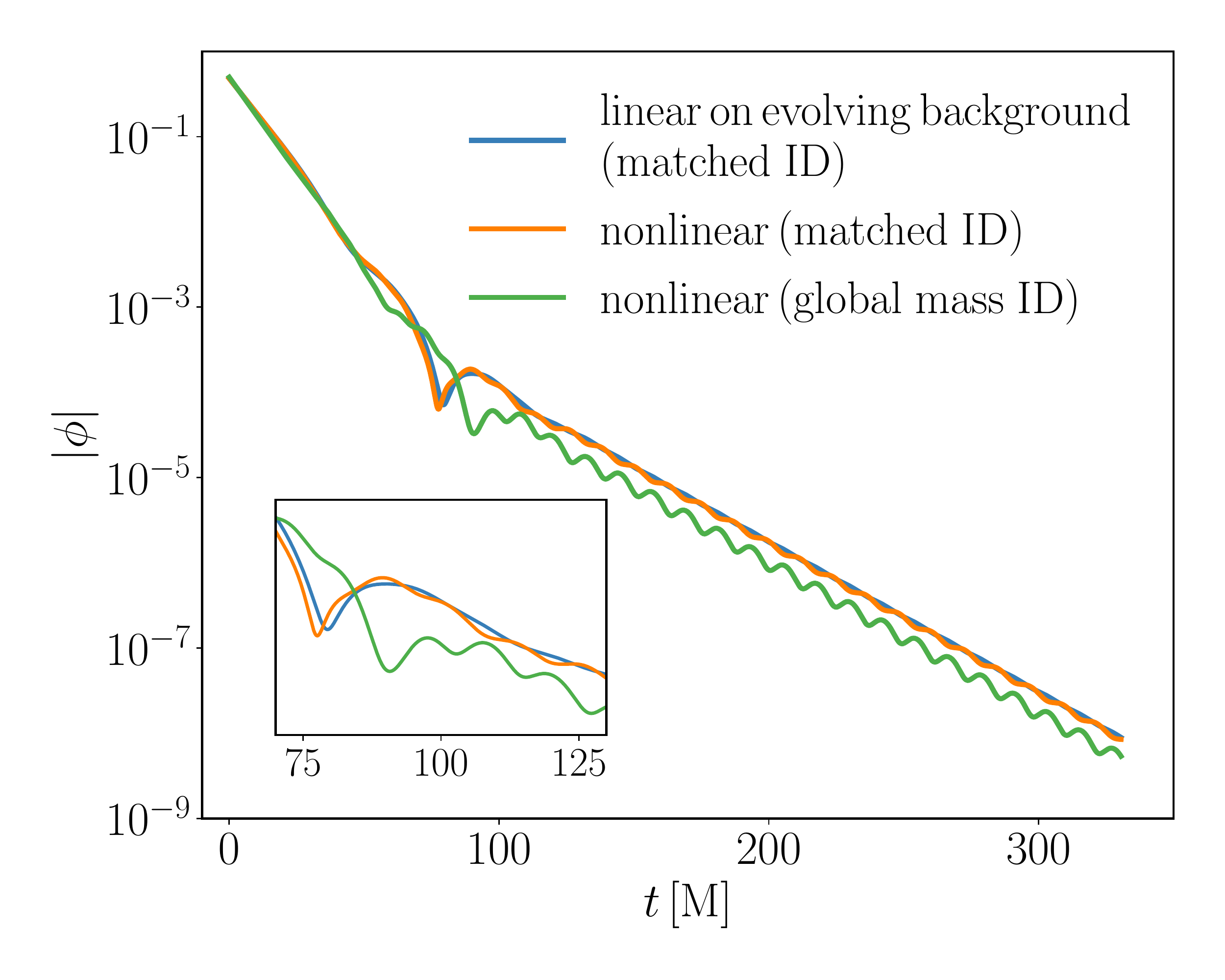}
  \caption{Fully nonlinear scalar field evolution compared to the linear
  evolution on a time-dependent black hole background. The tracks corresponding
  to two different choices of initial data are shown. They excite the
  fundamental modes approximately at the same level.  The inset shows in detail
  the transition between the overtone initial data and the nonlinearly excited
  fundamental mode that dominates the late-time evolution. Mode
  excitation is driven by absorption in both the linear and nonlinear
  evolutions.}
  \label{fig:linear_nonlinear}
\end{figure}

In Fig.~\ref{fig:linear_nonlinear}, we show the ringdown of the scalar
field in nonlinear simulations with two choices of scalar initial data:
the matched and the global-mass initial data defined in
Sec.~\ref{sec:numerical_fitting_setup}. Both simulations contain modes
excited nonlinearly and to a similar degree\footnote{Matched initial data
produces slightly stronger nonlinear excitations. This is because this
choice of initial data for the scalar field induces a smaller black hole
area on the initial slice, resulting in a larger area excursion to reach
the final value (set by the ADM mass, the same in all simulations). The
frequency of the initial perturbation, impacting the area evolution, is
also slightly different in the two choices of initial data.}. This tells
us that the excitations are mostly independent of the preparation of QNM
initial data. The mismatch between the background assumed for the initial
QNM profile and the actual background on the initial slice could have been
responsible for most of the nonlinear excitations--but this is not the
case.

We compare these nonlinear calculations in Fig.~\ref{fig:linear_nonlinear}
to the the result of a linear simulation where we evolve the scalar field
using the same initial data and the same parameters as used in the
(matched ID) nonlinear evolution, but on a SAdS background with a time
dependent mass set by the imaginary frequency of the initial data. 
We control the time dependence of the background black hole area as follows,
\begin{equation}
\mathcal{A}(t) = \mathcal{A}_{\rm init} + \Delta \mathcal{A} (1-e^{-t/\tau_{\rm back.}})
\label{eq:AnalyticAreaPrescription}
\end{equation}
where $\tau_{\rm back.}=2|\omega_{I, \hat{n}}|$
with $\omega_{I, \hat{n} }$ the imaginary frequency of the initial data
used, and $\Delta \mathcal{A} = \mathcal{A}_{\mathrm{fin}} -
\mathcal{A}_{\mathrm{init}}$. The three parameters $\omega_{I,\hat{n}}$,
$\mathcal{A}_{\mathrm{init}}$, and $\mathcal{A}_{\mathrm{fin}}$ are inputs of
the simulation and are chosen to match the nonlinear simulations. Specifically,
$\mathcal{A}_{\mathrm{fin}}=0.104$, and $\mathcal{A}_{\mathrm{init}}$ is the
mass used to compute the ID, which will be different depending on whether we use
the global-mass or matched ID.

In linear simulations, it is the evolution of the background alone
that produces the additional modes seen at late times in
Fig.~\ref{fig:linear_nonlinear}. This shows good agreement between
linear and nonlinear simulations with the same choice of scalar
initial data. This indicates that, among the possible sources
nonlinearities in our system, AIME is indeed the dominant one.

We also confirm that the excitations seen in the linear simulation are
related to the initial data through a cubic scaling (assuming the
black-hole mass change is dictated by the square of the perturbing field;
see Fig.~\ref{fig:amplitudes}). This shows that absorption excites
modes in the QN spectrum cubically, as argued in
Section~\ref{sec:QNMs_nonlinear}.


The linear setup, moreover, gives insight on the adiabaticity of
AIME in our system. We would like to understand if
the system is behaving adiabatically or nonadiabatically, and whether our
criterion based on the mode frequencies in Tab.~\ref{tb:table_AdStimes} is
reliable. To investigate this we evolve the $\hat{n}=1$ mode
in a background with a modified timescale for the black hole area change given by Eq.~\eqref{eq:AnalyticAreaPrescription} 
with $ \tau_{\rm back.} = \alpha \times \tau_{\rm back. \; true} = \alpha
\times ( 2 |\omega_{I, \; \hat{n} }|) $. 

In the nonadiabatic limit, $\alpha \rightarrow 0$, the background
instantaneously jumps to its final configuration, producing the largest excitations. In the opposite limit, $\alpha \rightarrow \infty$, the
background is frozen and no modes are excited that were not present in the
initial data; we observe the presence of the $\hat{n}=1$ overtone well into
intermediate times. For $\alpha =1$, we recover the true evolution induced by
the perturbing mode. 

\begin{figure*}[t]
	\centering
	\subfloat{
		\includegraphics[trim={1.cm 0.9cm 0.5cm 0.5cm},clip, width=0.49\linewidth]{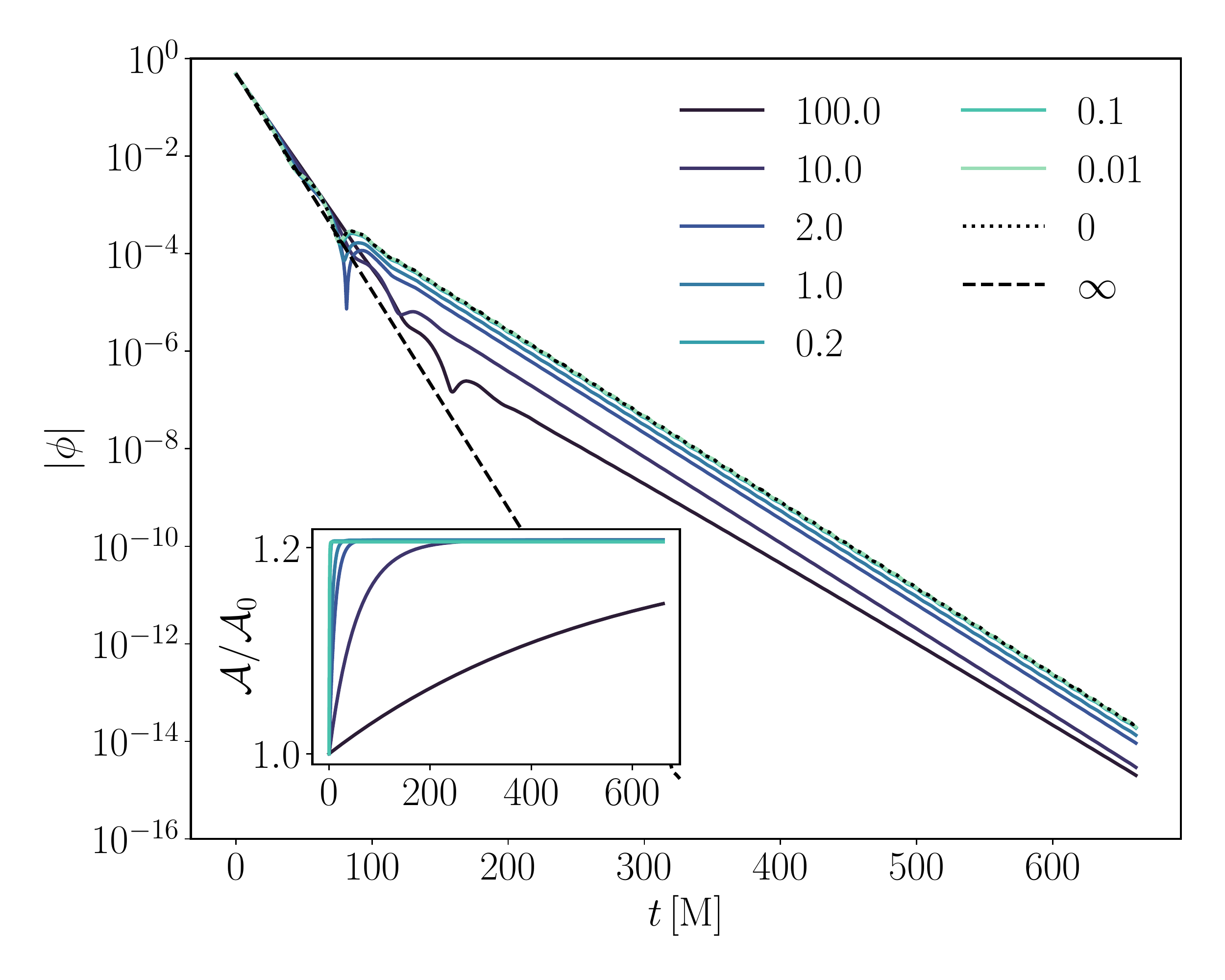} \hspace{0.1cm}
		\includegraphics[trim={.9cm 0.9cm 0.5cm .7cm},clip, width=0.49\linewidth]{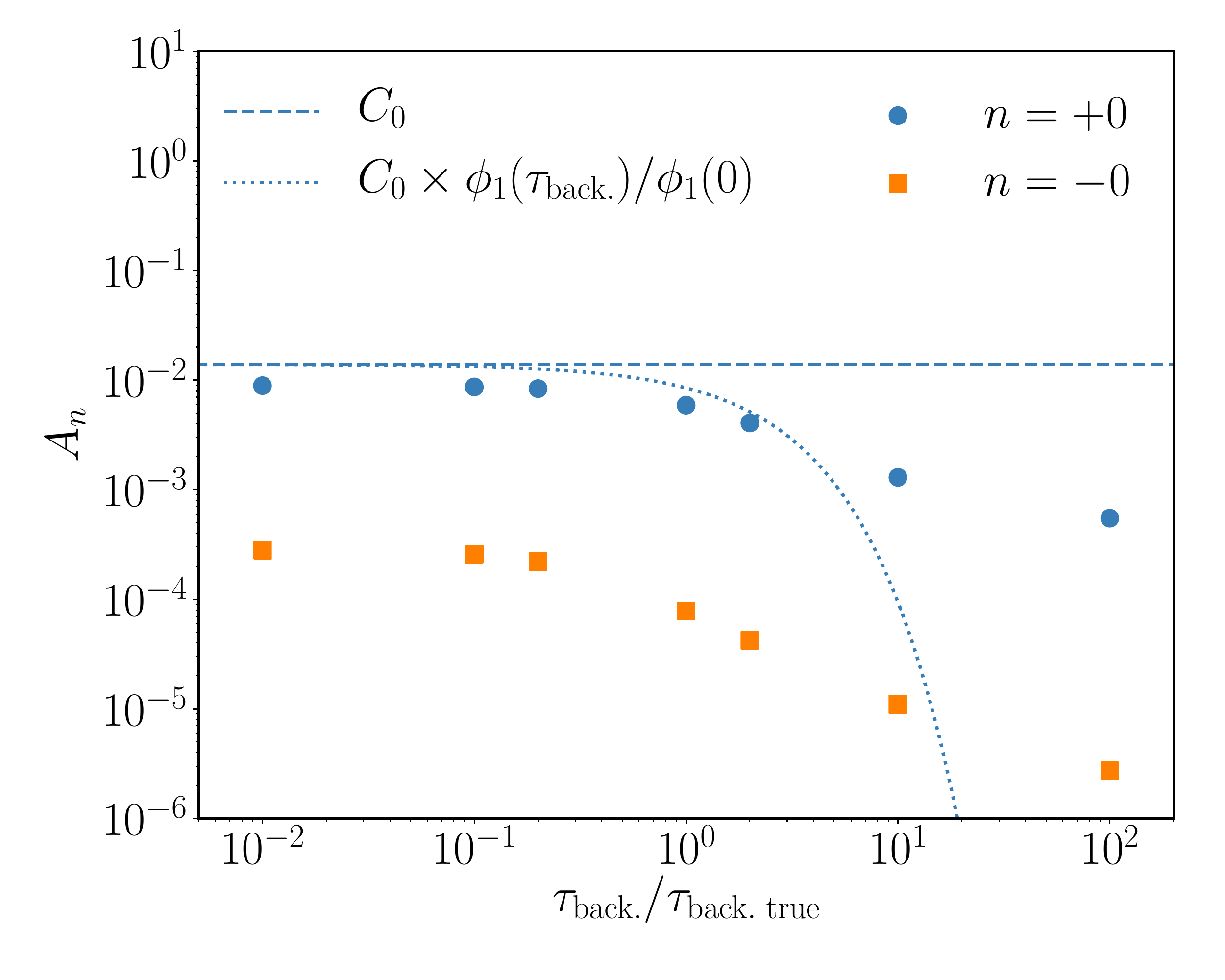} 
	}
	\caption{ \emph{Left}: Linear scalar field response to evolving background with evolution timescale $\tau_{\rm back.} = \alpha \; \tau_{\rm back. \; true}$. Different values of $\alpha$ (shades of blue) correspond to responses between two extremes: $\alpha=0$, where the background immediately jumps to a black hole mass inconsistent with the initial data, and $\alpha=\infty$, where the background is fixed and always consistent with the initial data. The evolution of the area of the black hole is shown in the inset (the time coordinate used at the AdS boundary is mapped to the Eddington-Finkelstein coordinate at the horizon via null rays).
		\emph{Right}: Amplitude of the nonlinearly excited fundamental modes as a function of $\alpha$. When the background evolves fast (small $\alpha$), the amplitude of the excitation is consistent with the nonadiabatic analytic estimate (dashed line). The nonadiabatic estimate can be improved at intermediate $\alpha$s accounting for the decay of the perturbing mode (dotted line). The initial data has $\hat{n}=1$, $A_1=0.1$. 
	} \label{fig:linear_adiab}
\end{figure*}

%
%
Figure~\ref{fig:linear_adiab} shows the resulting evolution of the scalar field for a range of values of $\alpha$, spanning the two extremes. 
Modes are excited with smaller amplitude when the background evolution is slower (large $\alpha$). In
those cases, we observe the mode's imaginary frequency shifting slowly in time.
The clearest example in Fig.~\ref{fig:linear_adiab} is for $\alpha=100$, the
slowest evolving background, where we observe the slope of the scalar field
amplitude settling into the final frequency, at late times. This slow evolution
of the scalar field amplitude is due to the time-dependence of imaginary
frequency induced by the slowly evolving background.  


The right panel of Fig.~\ref{fig:linear_adiab} shows how the excitation amplitudes converge to the nonadiabatic limit as $\alpha \rightarrow 0$. The true background evolution ($\alpha=1$) is itself close to the nonadiabatic limit, as its excitation amplitudes are only $\sim 30\%$ and $\sim 70\%$ smaller than the ones obtained with $\alpha =0.01$ for $n=+0$ and $n=-0$, respectively. We conclude that, when the background and mode timescale respect the condition $\tau_{\rm back.} \ll \tau_{\rm osc.}$, the nonadiabatic approximation can give a good order-of-magnitude estimate of the modes excited through this nonlinear process. 

Figure~\ref{fig:linear_adiab} also shows an analytic estimate of the
nonadiabatic excitation amplitude for $n=+0$. This estimate is obtained by
computing the excitation coefficient of the $n=+0$ mode of the late-time
black hole, with mass $M$. The initial data is set by the mode $\hat{n}=1$
of the initial black hole. We regularize the overlap integral between the
modes with the subtraction procedure described in
Sec.~\ref{sec:qnm_exc_linear}. The result is in good agreement (within
a factor of $2$) with the amplitude measured in the simulations in the
small $\alpha$ limit. We were not able to obtain a reliable estimate of
the excitation coefficient of the $n=-0$ mode, as our regularization
method has a precision comparable to the mode amplitude ($\sim 10^{-3}$).

The adiabatic limit, on the other hand, is harder to explore numerically. When
artificially slowed down, most of the evolution of the background takes place
when the amplitude of the perturbing mode has already decayed by orders of
magnitude. This is shown by comparing the area evolution in the inset and
the main plot in the left panel of Fig.~\ref{fig:linear_adiab}. In this regime, it is
hard to disentangle the effect of an adiabatic background evolution from the decay of the perturbation, as they both contribute to a reduced
excitation amplitude; when AIME becomes relevant,
the perturbation has already decayed a few orders of magnitude in amplitude. 
Indeed, the knee in the excitation coefficients as a function of the background timescale is consistent with the drop in the perturbing amplitude (right panel of Fig.~\ref{fig:linear_adiab}). At intermediate timescales, the decayed perturbation amplitude can be used to obtain a better analytic estimate of the nonlinear excitation, as shown in Fig.~\ref{fig:linear_adiab}. At very long timescales, on the other hand, the excitation amplitude does not fall as steeply as the perturbing mode. In this limit, the excitation could be explained by the partial area increase taking place before the complete decay of the perturbing mode.



\section{Towards astrophysical black holes}\label{sec:asymp_flat}
\begin{figure}[t]
	\centering
	\includegraphics[trim={.5cm 0.1cm 2.5cm 2.cm},clip,width=0.96\linewidth]{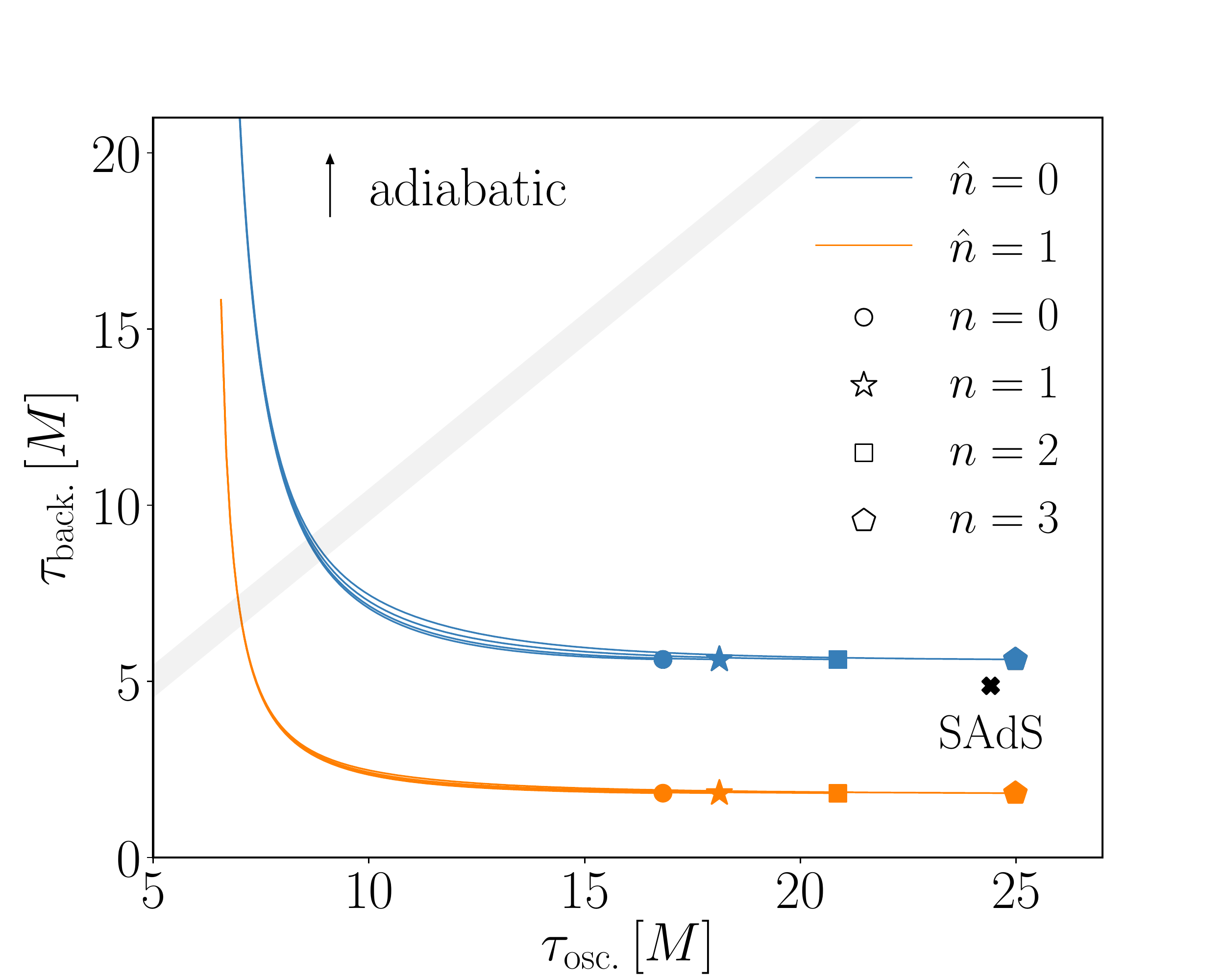}
	\caption{Timescale of mode oscillation $\tau_{\rm osc.}=2 \pi /\omega_{R,n}$ vs.~timescale of background evolution driven by $\hat{n} =0 $, $\tau_{\rm back.}=1 / 2\omega_{I, \; \hat{n}}$. We consider overtones of Kerr with spin ranging in $a= [0,0.999]$ (Schwarzschild is denoted by a marker) and $l=m=2$. The gray line approximately separates the adiabatic regime, where $\tau_{\rm back.} \gg \tau_{\rm osc.}$, from the complementary non adiabatic regime. In the AdS example from previous sections, the oscillation timescale is set by the fundamental mode, while the backreaction timescale is set by the first overtone. 
	} \label{fig:timescales_Kerr}
\end{figure}

In this section, we connect our results to gravitational-wave
observations of ringdowns. Although so far we have considered scalar
perturbations of asymptotically AdS black holes, the generic effect of
overtones being excited due to an evolving black-hole background will
occur just as well for gravitational perturbations of asymptotically
flat black holes, e.g., in the ringdown of a merger remnant. Although
a full analysis of this case is beyond the scope of this paper, here,
building on the success of the analytic predictions in reproducing the
nonlinear excitations seen in SAdS, we make similar predictions for
\emph{gravitational} perturbations of \emph{Schwarzschild} black
holes.  We estimate the effect of AIME using the same methodology as
described in Sec.~\ref{sec:theorybeyond}. After summarizing the
theoretical framework, we evaluate the excitation coefficients
numerically to estimate the significance of the backreaction
phenomenon for gravitational wave observations.  Though here we
restrict our analysis to non-spinning black holes for simplicity, we
also outline the steps required to extend our results to Kerr.

At second order in perturbation theory, vacuum gravitational perturbations of
Kerr were computed numerically for the first time in
Ref.~\cite{Ripley:2020xby}. At this order, perturbations are excited through
mode mixing, so that a linear perturbation of azimuthal number $m= \pm \hat{m}$
can only excite modes with $m= \pm 2 \hat{m}$ and $ 0 $. Our focus is on
nonlinear excitations mediated by the black hole area evolution, which first appears at
\emph{third order} in perturbation theory. Na\"{i}vely, its perturbative order suggests that AIME might be subdominant. This might not be the case, as we discuss in Section~\ref{sec:discussion}.

We work again in the horizon penetrating, ingoing Eddington-Finkelstein coordinates, Eq.~\eqref{eq:metricEF} with $ f(r) = 1- 2 M/r  $. The tortoise coordinate is defined as $ dr_*/dr = 1/f(r) $ and the horizon is located at $ r_+=2 M $. 

\subsection{QNM horizon flux and area evolution}

We begin by deriving the evolution of the Schwarzschild black hole area induced by the QNM flux. We follow closely the derivation of Teukolsky and Press~\cite{TeuPress}, adapting it to complex frequencies.

The area of a Schwarzschild black hole is given by
\begin{equation}\label{key}
\mathcal{A} = 16 \pi M^2.
\end{equation}
Thus, the evolution of the mass and the area are related by
\begin{equation}\label{key}
\frac{d^2 \mathcal{A}}{dv d\Omega} = 32 \pi M \frac{d^2 M}{dv d\Omega} . 
\end{equation}
The energy flux can be expressed in terms of the Weyl scalar $\psi_0$, giving
\begin{equation}\label{key}
\frac{d^2 \mathcal{A}}{dv d\Omega} = 32 f^4 M^5 \frac{|\psi_0|^2}{|1+i 4 M \omega|^2}.
\end{equation}
Taking the horizon limit,
\begin{equation}\label{key}
\psi_0 \sim e^{-i \omega v + i m \phi} {}_2S_{lm}(\theta) \psi_{0,r_+} r_+^{-4} f^{-2},
\end{equation}
we find
\begin{equation}\label{eq:SchwdA}
\frac{d^2 \mathcal{A}}{dv d\Omega} =  \frac{|\psi_{0,r_+}|^2 }{8 M^3 |1+i 4 M \omega|^2} |{}_2S_{lm}(\theta)|^2 e^{2 \omega_I v} . 
\end{equation}
This is more conveniently expressed in terms of the amplitude of the Weyl scalar $\psi_4$,
\begin{equation}\label{eq:psi4_hor}
 \psi_4 \sim \psi_{4,r_+} e^{-i \omega v + i m \phi} {}_{-2}S_{lm}(\theta)   f^{2}, \quad r \rightarrow r_+ .
\end{equation}
The Teukolsky-Starobinski identities relate the amplitudes of $\psi_4$ and $\psi_0$ via the Starobinsky constant $C$,
\begin{equation}\label{key}
C \psi_{0,r_+} = 512 M^5 i \omega |1- 4 i M \omega|^2 (1- 2 i M \omega) \psi_{4, r_+}.
\end{equation} 
The Starobinsky constant for complex frequencies is obtained via the analytic continuation
\begin{equation}\label{key}
C = E(E-2) +  12 i M \omega ,
\end{equation}
where\footnote{Note that the eigenvalue $E$ is independent of the sign of $s$ in the absence of the black hole spin.} $ E=E^{|s|}_{lm} $ is the eigenvalue of the spin-weighted spherical harmonics ${}_s S_{lm}$, normalized as $\int d\theta |{}_s S_{lm}(\theta)|^2 \sin \theta = 1$. The amplitude of the Weyl scalar $\psi_{4,r_+}$ at the horizon can be related to the amplitude at infinity $\psi_{4,\infty}$ through the global solution, which we evaluate numerically. Finally, $\psi_{4,\infty}$ is related to the two gravitational wave polarizations at infinity $\psi_{4}(r\rightarrow \infty) = \frac{1}{2} ( \ddot{h}_+-i \ddot{h}_{\times}) $ through the asymptotic limit, 
\begin{equation}\label{eq:psi4_inf}
 \psi_4 \sim \psi_{4,\infty} e^{-i \omega v + i m \phi}{}_{-2}S_{lm}(\theta)   r^{-1}, \quad r \rightarrow \infty .
\end{equation}
Dots denote derivatives with the respect to the Boyer-Lindquist time $t$.

\subsection{Nonlinear excitations due to absorption}
The nonlinear excitation of modes due to absorption
depends, as we argued in the SAdS case, on how fast backreaction takes
place compared to the mode oscillation. Figure \ref{fig:timescales_Kerr}
shows that for the $l=m=2$ mode, which dominates most observations of
ringdowns, the fundamental and first overtone can both give rise to
nonadiabatic background evolution for spins $a \lesssim 0.9$. In
particular, the AIME excitation of the $n=3$ mode driven by the
absorption of the fundamental mode should be very similar to the SAdS
example discussed in the previous section. Thus, from the point of
view of modes in Schwarzschild, backreaction happens almost
instantaneously. This justifies using an instantaneous projection to
estimate nonlinear excitation amplitudes, as done in
Sec.~\ref{sec:linear_sim}. In the following, we summarize the theory
of the Schwarzschild excitation coefficients, specializing to the case
of QNM initial data.

Given initial data in Schwarzschild, the excitation coefficient of a QNM can be computed in the Laplace-transform formalism~\cite{Berti_2006}. The Laplace transform of the radial Teukolsky equation reads
\begin{equation}\label{eq:LaplaceTeu}
\mathcal{O}^s_{lm} \tilde{R}^s_{lm}(r) = \mathcal{S}^s_{lm},
\end{equation}
where the radial function is defined via the separation of the Teukolsky perturbation variable
\begin{align}\label{key}
\Upsilon_s(\omega,r) =&R^s_{lm}(\omega,r) e^{im\phi} {}_s S_{lm}(\theta) \nonumber  \\ =& \frac{1}{\sqrt{r^2r^{2s}f^s}} \tilde{R}^s_{lm}(\omega,r) e^{im\phi} {}_s S_{lm}(\theta),
\end{align}
and where 
\begin{align}\label{key}
\mathcal{O}^s_{lm} =& \partial^2_{r_*} +  V^s_{lm}(r),\\
V^s_{lm}(r)=& \omega^2 -\frac{f^2 \left(s^2-4\right)+s (s+4 i r \omega )}{4 r^2}, \nonumber \\
& + \frac{f}{2 r^2} \left(2 \lambda -6 i r s \omega +s^2+2 s+2\right),
\end{align}
with $ \lambda =E-s(s+1)$.

We assume that the initial data is a QNM with frequency 
$\hat{\omega} = \omega_{\hat{n},\hat{l},\hat{m}} $, so that
$ R^{s }_{lm} (t=0)= R(\hat{\omega}) $ and
$ \dot{R}^{s}_{lm}(t=0) = -i \hat{\omega} R(\hat{\omega} ) $, where we
suppressed the QNM indexes. Then, the source in
Eq.~\eqref{eq:LaplaceTeu} reads~\cite{Campanelli:1997un}
\begin{align}\label{key}
\mathcal{S}^s_{lm}=\mathcal{S}(\omega, \hat{\omega} ) = f^{\frac{s}{2}} r^{s} \bigg[ 
i (\hat{\omega}  +\omega) r + s -3 s f
\bigg] R(\hat{\omega} ).
\end{align}
%
The excitation coefficient for a mode of frequency $\omega$  is defined as
\begin{equation}\label{key}
C_{\omega}(\hat{\omega}) = B_\omega \, \mathcal{O}(\omega,\hat{\omega} ),
\end{equation}
where we defined the overlap integral
\begin{equation}\label{key}
\mathcal{O}(\omega,\hat{\omega} ) = \int_{\mathcal{C}} dr_* \mathcal{S}(\omega, \hat{\omega} )  \tilde{R} (\omega) .
\end{equation}
The integral is defined on a complex deformation of the real line, $\mathcal{C}$, so as to be convergent in the two limits $r_{*}\rightarrow \pm \infty$. Based on the asymptotic form of the QN radial functions at infinity and at the horizon, the contour is given by \cite{Green:2019oaxaca,Stephen} 
\begin{equation}
	r_* \rightarrow \Lambda + \rho e^{i \theta}, \quad \theta \in [-\pi - \arg(\omega+\hat{\omega} ), - \arg(\omega+\hat{\omega} )],
	\end{equation}
for $r_*>\Lambda>0$, and 
\begin{equation}
	r_* \rightarrow \lambda + \rho e^{i \theta}, \quad \theta \in [- \arg(\omega+\hat{\omega} ), \pi - \arg(\omega+\hat{\omega} )],
\end{equation}
for $r_*<\lambda<0$, with $ \rho>0 $ in the two regions. The resulting integral, which is the sum of three contributions ($r_*<\lambda$, $r_*>\Lambda$, and the intermediate integral on the real line) is independent of $\Lambda$, $\lambda$, and the choice of $\theta$ within its limits. 

We adopt the following definition for the excitation
factors~\cite{Green:2019oaxaca,Stephen}
\begin{equation}\label{key}
B_\omega = \left[ \int_{\mathcal{C}} dr_* \mathcal{S}(\omega, \omega)  \tilde{R} (\omega)  \right]^{-1} ,
\end{equation}
which guarantees the overall normalization of the excitation coefficients. The excitation factors of a Schwarzschild black hole can also be computed from the asymptotic expansion of the solutions of the Teukolsky equation, see for instance Refs.~\cite{Zhang_2013,Oshita:2021iyn}.

\subsection{Numerical estimates}

%
As an example, we consider an initial perturbation composed of a single
fundamental ($\hat{n}=0$) mode with\footnote{We choose a
mode with vanishing $m$ because it does not backreact on the black hole
spin, and we can consistently study the problem in a Schwarzschild
background.} $ l=2$, $m=0 $, and estimate the nonlinear excitation of overtone modes with
the same angular numbers under the approximation that the mass evolves
instantaneously. The timescales for these modes are shown in
Fig.~\ref{fig:timescales_Kerr} (in Schwarzschild, mode frequencies are
independent of $m$).

We find the QNFs and the radial functions for $s=-2$ using the Black
Hole Perturbation Toolkit~\cite{BHPToolkit} for a given initial black
hole mass $M_{\rm init}$.  For initial data consisting of a single (the fundamental)
QNM, we then estimate the final mass of the black hole due to the
inward flux using Eqs.~\eqref{eq:SchwdA}--\eqref{eq:psi4_inf}.  We
then re-expand the initial mode with
$ \hat{\omega} =\omega_{0}(M_{\rm init}) $ in terms of the overtones
of the final black hole, $ \omega =\omega_{n}(M_{\rm final}) $. We
verified that the overlap integral has the expected scaling with the
mass variation, $ \mathcal{O} \sim \Delta M $ for
$\Delta M \ll M_{\rm init}$, by evaluating it numerically for a range
of mass increments; see Appendix~\ref{app:convergence}.
%
\begin{table}[t]
		\begin{ruledtabular}
			\begin{tabular}{ccccc}
		$n$=0 (pert.) 
		&$n$=0 (exc.)& $n$=1
		& $n$=2
		& $n$=3	\\
		\colrule
		$1\times 10^{-2}$&$2\times 10^{-4}$&$6\times 10^{-5}$&$8\times 10^{-5}$&$9\times 10^{-5}$\\ 
		$5\times 10^{-2}$&
		$1.4\times 10^{-2}$
		&$7\times 10^{-3}$&$8\times 10^{-3}$&$8\times 10^{-3}$\\
	\end{tabular}
\end{ruledtabular}
	\caption{Nonadiabatic excitation coefficients up to the third overtone, produced by fundamental mode ($n$=0) initial data. The initial amplitude of the fundamental mode is given in the first column. The initial perturbation produces a variation of the mass of $\Delta M/M= 10^{-3}$ and $\Delta M/M \simeq  2.6 \times 10^{-2}$, respectively. The excitation coefficients refer to the amplitude $|\psi_{4,\infty}|$. }\label{tb:table_exc_Schw}
\end{table}
\begin{figure}[th!]
  \centering
  \includegraphics[trim={0.cm 1.5cm 1.5cm 3.5cm},clip,width=0.96\linewidth]{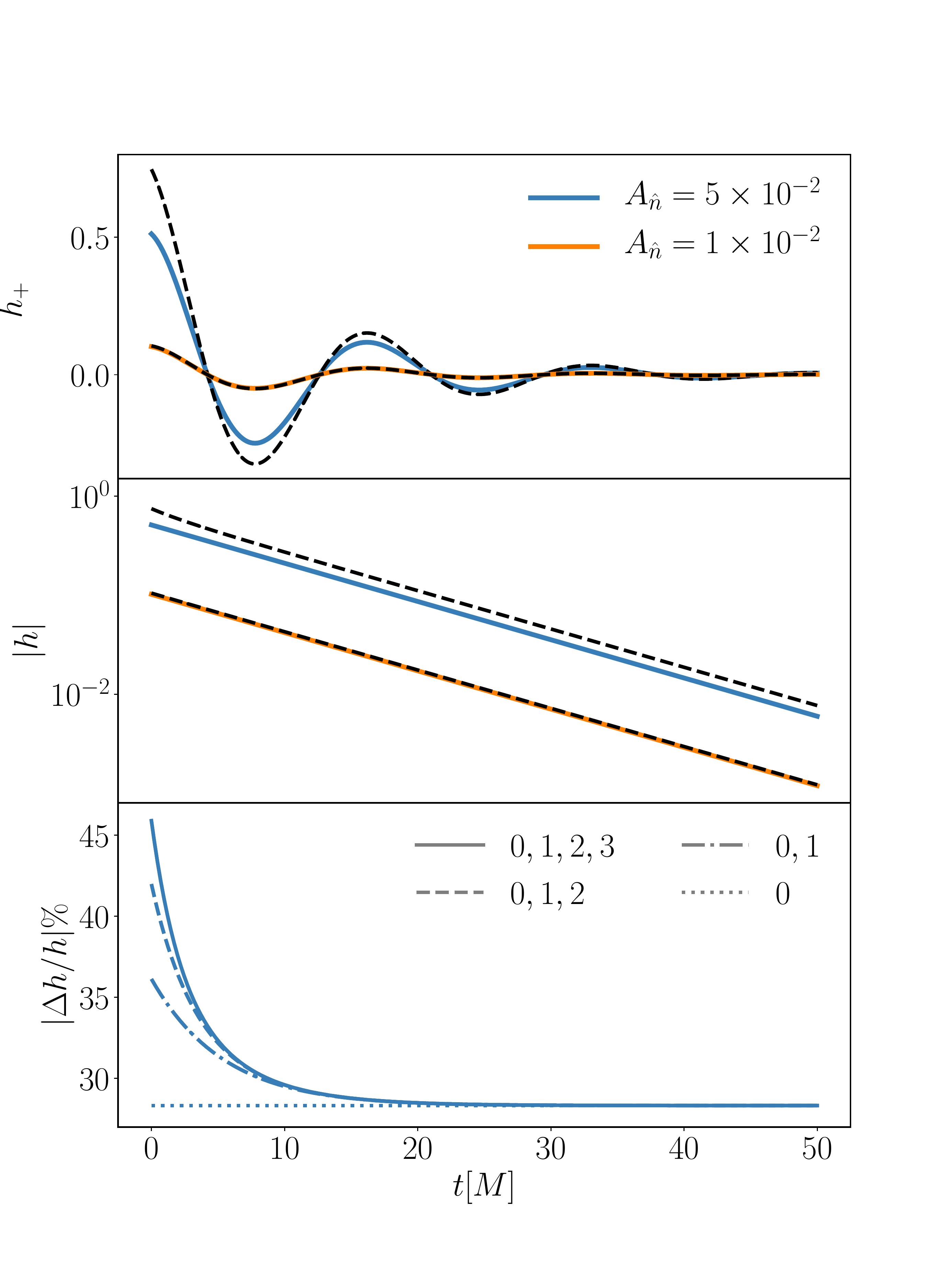}
  \caption{ The linear (solid lines) and the estimated nonlinear
    ringdown (dashed) for perturbations with two different amplitudes
    and $\hat{n}=0$, $l=2$, $m=0$. To construct the nonlinear signal,
    we use the nonadiabatic excitation coefficients in a Schwarzschild
    background, (Table \ref{tb:table_exc_Schw}) and only include
    overtones up to $n=3$ with the same $(l,m)$ as the linear
    perturbation. The effect of the nonlinear excitation is
    significant (percent level or more) in both examples. (For
    $A_n=1\times10^{-2}$, the residual is not shown in the bottom
    panel to improve legibility; it asymptotes to $2 \%$.) The
    amplitude of the gravitational wave of the single mode is related
    to the Weyl scalar amplitude reported in
    Table~\ref{tb:table_exc_Schw} by
    $h_n (t) \equiv h_n(t,r)\; r/M \sim 2 \, \psi_{4,\infty,n} /
    \omega^2 e^{-i\omega_n t}$ for $r \rightarrow \infty$.
  } \label{fig:excitation_gw}
\end{figure}

The excitation coefficients of the first three overtones are given in
Tab.~\ref{tb:table_exc_Schw}, for two values of the initial perturbation
amplitude. For simplicity, we only consider modes with $\omega_R>0$ and neglect the decay of the perturbing mode, which (as seen in Sec.~\ref{sec:linear_sim} and Fig.~\ref{fig:linear_adiab}) can slightly suppress the excitations. We find that, for
realistic values of the linear perturbation amplitude,\footnote{In Ref.~\cite{Giesler:2019uxc}, the fundamental mode in the numerical relativity simulation resembling GW150914 is found to have amplitude of order one. The perturbations in our two examples correspond to gravitational wave amplitudes of $0.1$--$0.5$, see Fig.~\ref{fig:excitation_gw}.} nonlinear
excitations can affect the ringdown signal at the percent level. The first three overtones have comparable excitation amplitudes, with the latter growing slightly with $n$. Indeed, their excitation \emph{factors}, the measure of their intrinsic excitability, grow with $n$ peaking around $n=4$ to $6$ for nonextremal black holes, as recently shown in Ref.~\cite{Oshita:2021iyn}. 

To visualize
the effect of these nonlinearities, we show the corresponding total
gravitational wave at infinity in Fig.~\ref{fig:excitation_gw}. For simplicity, we assume that all modes are sourced with the same (zero) phase. The impact of AIME on the gravitational wave signal is significant: it is above the percent level for both choices of the perturbation amplitude. The nonlinear excitation in the fundamental mode has the strongest impact, both at early and late times. The effect of the nonlinearly excited overtones is also significant at early times. 

Following a black hole binary inspiral, the merger itself
may seed overtones. In a realistic scenario, therefore, the nonlinearly generated modes
will add to those sourced directly.
Furthermore, we emphasize that in producing Fig.~\ref{fig:excitation_gw}, for simplicity, we
did not take into account any delay between the linear and the nonlinear
excitation. The precise onset of nonlinear excitations requires
further investigation, as discussed for SAdS.



\subsection{The spinning case: expectations}

In this work, we studied nonlinearities in the ringdown of nonspinning
(Schwarzschild and Schwarzschild-AdS) black holes. The generic merger
of an astrophysical black hole binary will, however, result in a
spinning (Kerr) black hole. It is therefore essential to extend the
study of nonlinear effects to the spinning case to make better contact
with gravitational wave observations.  A detailed analysis of the Kerr
ringdown beyond linear order is left for future work. Here we
summarize the main conceptual differences compared to the nonspinning
case.

Backreaction onto a spinning black hole will change its angular
momentum in addition to its mass, according to
\begin{equation}\label{eq:dJ}
dJ = \frac{m \, \omega_{\rm R}}{|\omega|^2} dM
\end{equation}
for a single mode; see Appendix~\ref{app:dJdM}. As a consequence, a linear
perturbation will need to be re-expanded not only in terms of modes
with different black hole mass, but also with different spin. Since the
angular solution (a spin-weighted spheroidal harmonic) now depends on
the spin, different $l$ modes would generically be excited through the
nonlinear process. Modes with different $m$ will remain
orthogonal. Note that the excitation amplitude will depend on the
perturbing mode number $m$ through the change in the black hole spin,
as predicted by Eq.~\eqref{eq:dJ}.

%
\section{Discussion and conclusions}\label{sec:discussion}

The final phase of a black hole binary merger, the ringdown, is an ideal testing ground for general relativity. Some time following the merger, the signal should be well described by linear perturbations---in particular, QNMs---of an isolated black hole with fixed mass and spin. The goal of the black hole spectroscopy program is to identify these modes and accurately measure their frequencies in gravitational wave data.

Our goal in this work was to explore the dynamics of perturbed black holes
beyond linear order. With numerical simulations of a toy model (a scalar field
coupled to gravity in SAdS), we elucidated an important nonlinear effect,
entering at higher order in perturbation theory and with secular characteristics. 
We verified that the flux of QNMs through the horizon can change the black hole mass during the early ringdown.
This is a second order effect on the metric, which affects the QNFs (which depend on the background). 
The evolution of the black hole can also excite QNMs at third order in perturbation theory. We call this effect AIME. Although for gravitational waves (in contrast to a scalar field),
nonlinear effects start at second order (e.g., mode doubling),
the mechanism described here secularly accumulates as the initial modes decay,
and can become comparatively significant. 

Through detailed analysis in asymptotically AdS black-hole contexts,
we find that AIME can be captured with fully nonlinear calculations,
as well as through linear simulations on a time-dependent
background. We also provide a simple method to estimate the amplitude
of the excited modes, based on QN excitation coefficients.  This
method hinges on the fact that the background evolves
non-adiabatically compared to the mode oscillations. We expect that
the intuition developed for the toy model carries over to
asymptotically flat spacetimes. We estimate the amplitude of AIME in
Schwarzschild and show that this effect could make a percent-level
contribution to the ringdown following a merger.

Our analysis has implications for ringdown studies of generic black
hole merger gravitational wave events:
\begin{itemize}
	\item Overtones are excited generically and dynamically in the ringdown at higher orders in perturbation theory. 
	
	\item The mass (and spin) of the remnant black hole necessarily evolve over time in the aftermath of the merger. Ringdown analysis looking for QNMs near the peak of the signal should take into account that their frequencies can evolve with time. In particular, inference based on signals detected early in the ringdown,
while showing a higher number of modes, will bias the extracted value towards a smaller black hole mass.
Later on, longer-lasting lower modes will allow for a more accurate measurement of the final mass.
	For this reason, higher overtones could actually bias rather than improve the inference of the (final) remnant properties, as seen in some numerical simulations~\cite{Finch:2021iip,Li:2021wgz}.
	\item The amplitude of the QNMs should evolve also over time in the early ringdown as a result of AIME. The evolution should happen on the same timescale as the black hole mass and spin. 
	\item Overall, our results suggest that ringdown analysis---especially when starting early
	after the merger---could benefit from being made 
	more flexible, allowing for the evolution of mode frequencies and amplitudes.  	
\end{itemize} 

Beyond these practical observations, the larger question of nonlinear
behavior in the post-merger regime is interesting in its own right. At
second order in perturbation theory, a known effect is the energy
transfer between modes with different azimuthal number $m$, explored
in Ref.~\cite{Ripley:2020xby}. Despite appearing at higher order in
perturbation theory, AIME could be as significant to the ringdown as
second order mixing. Although a direct comparison is not
straightforward, the excitations of the Weyl scalar measured at null
infinity in \cite{Ripley:2020xby} are of the same order
($10^{-4}-10^{-5}$ for a $10^{-2}$ perturbation) as the analytic
estimates reported in our Table~\ref{tb:table_exc_Schw}. Indeed, AIME
can be thought of as a secular effect, due to the cumulative
change in the black hole mass, in contrast to mode mixing, whose
source oscillates and strictly decreases as the linear mode
decays. This may upend the order counting argument. Second order
calculations might also be affected by the secular mass evolution
discussed in this work, depending on the stage at which they are
carried out. Finally, our mechanism can excite all the overtones of
the initial perturbation of azimuthal number $\hat{m}$, and can
therefore be complementary to mode-mixing excitations of different $m$
modes. Overtones of the $l=|m|=2$ mode are already being used in
analyzing the ringdown of gravitational wave events to extract the
mass and spin of the remnant~\cite{Isi:2019aib}, while higher angular
modes in the ringdown have proven more elusive~\cite{Capano:2021etf}
(see however, \cite{cotestainprep}).

While our work elucidates nonlinear mode excitations in perturbed black holes in general settings
through lessons drawn in asymptotically AdS scenarios,  
further studies in asymptotically flat spacetimes would be valuable.
In particular, linear simulations on a time-dependent
background, as applied in this work to asymptotically AdS black holes, could be
easily used to further explore AIME in asymptotically flat black holes. The problem becomes richer in Kerr, where modes with different angular number $l$ could be excited as well. 
Particularly intriguing would be exploring our
mechanism for mode excitation in the extremal limit of Kerr, where QN
frequencies seem to suggest that the adiabatic limit might apply
(Fig.~\ref{fig:timescales_Kerr}) so AIME would be suppressed. AIME should also apply in the context of superradiance, where modes, although not absorbed, nevertheless modify the background spacetime. Finally, modes should be similarly excited when black holes accrete matter.


Finally, our work shows that black holes do not act like censors of 
nonlinearities after merger.  On the contrary, by efficiently
absorbing energy and growing, they mediate the nonlinear
AIME.  


\begin{acknowledgments}
We thank C.~Kavanagh, S.~Hollands, N.~Turok, N.~Warburton, B.~Wardell, and P.~Zimmerman for
useful discussions.  This work makes use of the Black Hole Perturbation
Toolkit.  W.E. and L.L. acknowledge support from an NSERC Discovery grant.
L.L. acknowledges CIFAR for support.  Research at Perimeter Institute is
supported in part by the Government of Canada through the Department of
Innovation, Science and Economic Development Canada, and by the Province of
Ontario through the Ministry of Colleges and Universities.
\end{acknowledgments}

\appendix
\section{Orthogonality of the QNMs}\label{app:convergence}
We confirmed the orthogonality of modes with different overtone numbers on a given background spacetime. 

\begin{figure}[t!]
	\centering
	\includegraphics[trim={0.2cm .cm 1.cm 1.2cm},clip,width=0.96\linewidth]{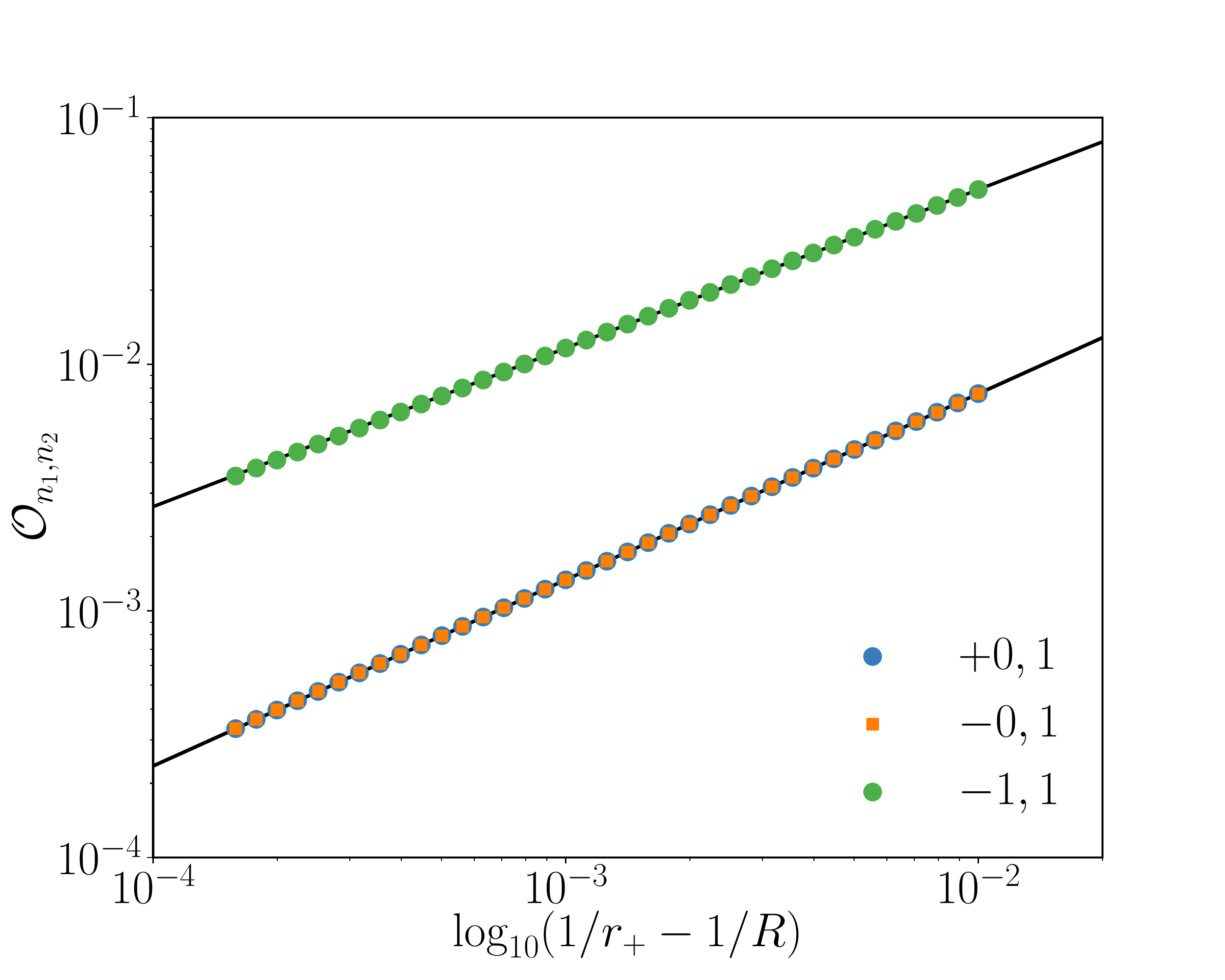} 
	\caption{ Overlap integral between scalar QNMs. The background is SAdS with black hole mass $M=0.104$. As the regulator $R$ is taken closer to the horizon, the overlap between modes with different overtone number goes to zero. Black lines are power-law fits, $ \sim a  x^b$; we find $b>0$ for all combinations of modes.
	} \label{fig:convergence_ortho_overlap}
\end{figure}

Figure~\ref{fig:convergence_ortho_overlap} shows the case of scalar modes in SAdS. The orthogonality is achieved as the regulator $r=R$ is taken to the black hole horizon. This is equivalent to the limit $R_*(R)\rightarrow-\infty$ used in the main text, Section~\ref{sec:qnm_exc_linear}.  

\begin{figure}[t]
	\centering
	\includegraphics[trim={0.cm 0.1cm 1.3cm 1.3cm},clip,width=0.96\linewidth]{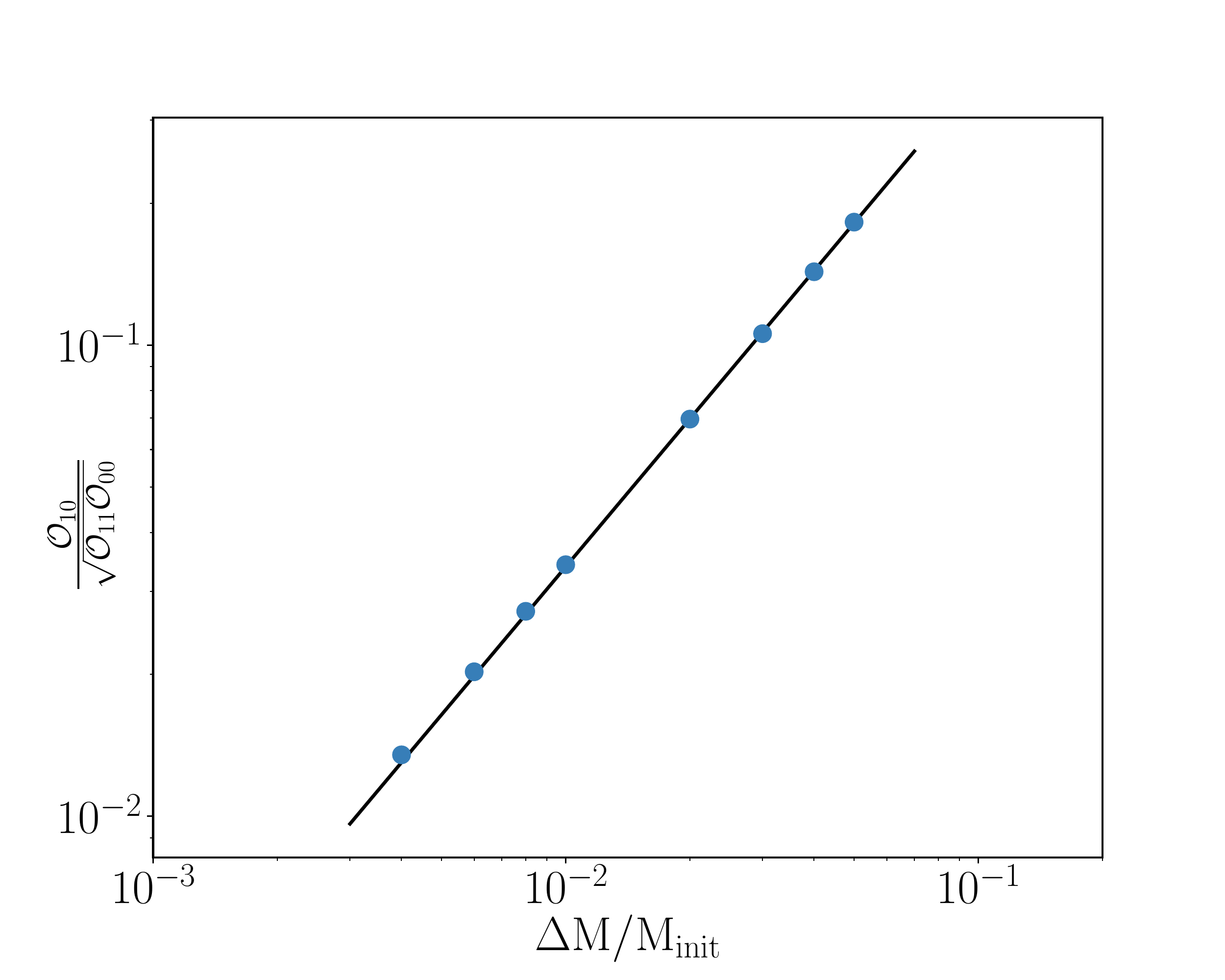}
	\caption{Overlap integral of the first overtone in a Schwarzschild background with mass $M_{\rm init}+\Delta M$, sourced by a fundamental mode corresponding to a black hole mass $M_{\rm init}$. The label should be read as $\mathcal{O}_{1 0} = \mathcal{O}(\omega_1(M_{\rm final}),\omega_{0}(M_{\rm init})) $ and $\mathcal{O}_{1 1 (00)} = \mathcal{O}(\omega_{1 (0)}(M_{\rm init}),\omega_{1 (0)}(M_{\rm init})) $. 
		The solid line is a fit to the numerical overlaps, $\mathcal{O} \sim \Delta M^{1.04}$. 
	} \label{fig:overlap_scaling}
\end{figure}

We obtain a similar result, with a different regularization procedure, for gravitational modes of Schwarzschild. In both cases, we checked that the overlap integral had the expected scaling with the perturbation amplitude, when computing the overlap between the perturbation and the mode living in the perturbed spacetime. This is shown in Fig.~\ref{fig:overlap_scaling} for Schwarzschild.

\section{Goodness of the fits}\label{app:fits}
To assess the goodness of the fit, we estimate the numerical noise at each time $ t $ as the residuals between our main datasets and higher resolution simulations (twice the grid-size) times a second-order Richardson extrapolation factor: $ 4/3 \times |\phi-\phi^{\text{ higher res.}} |$. The global goodness of the fit is determined by computing the mismatch,
\begin{equation}\label{eq:mism}
\mathcal{M} = 1 - \frac{\langle \phi ,\phi^{\rm fit} \rangle}{\sqrt{\langle \phi ,\phi \rangle \langle \phi^{\rm fit} ,\phi^{\rm fit} \rangle }} \, ,
\end{equation}
where $ \langle f, g \rangle = \int_{t_0}^T f(t) g^*(t)\, dt $.

\begin{figure}[ht!]
	\centering
	\includegraphics[trim={0.9cm 1.cm .5cm 0.9cm},clip,width=0.96\linewidth]{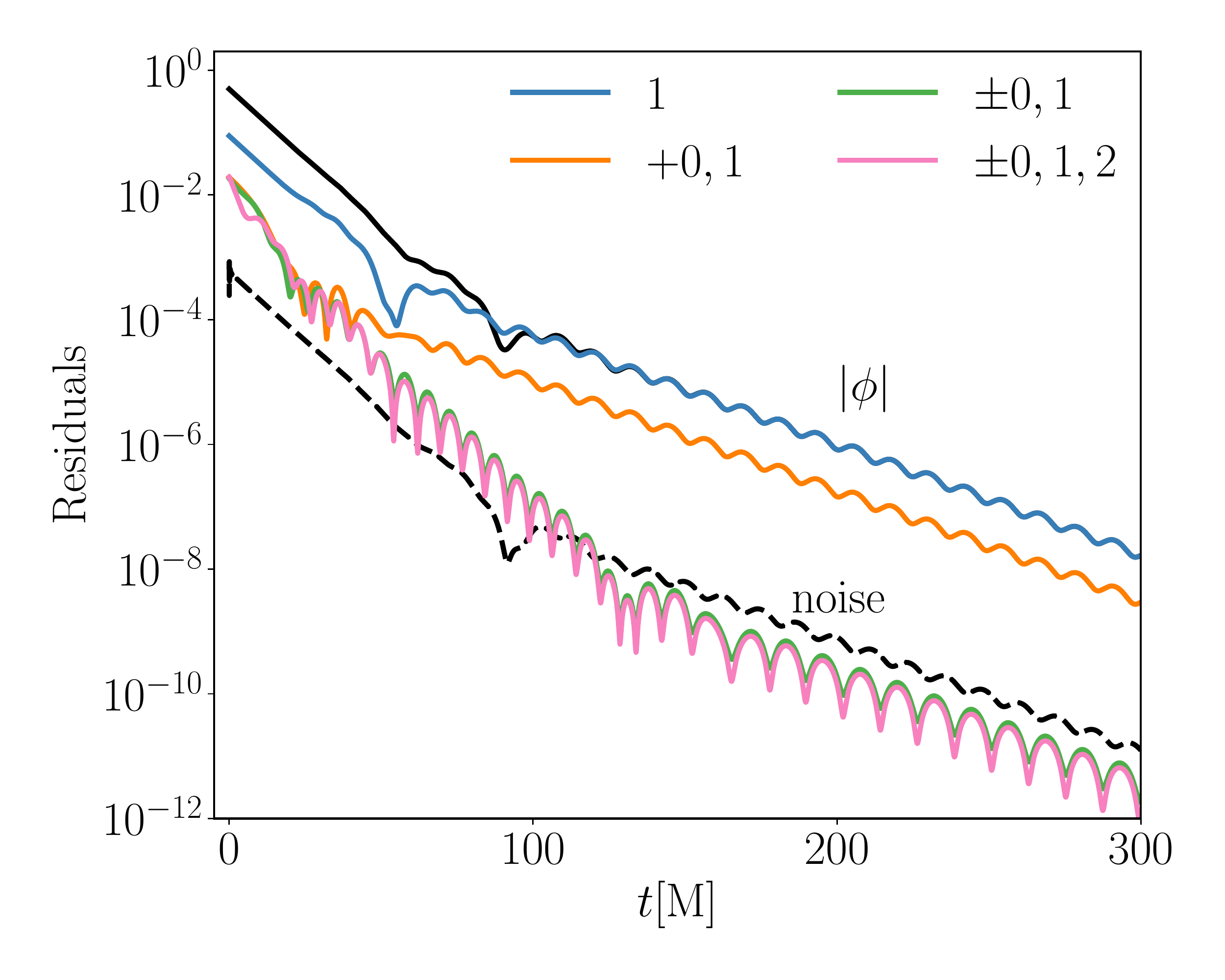} 
	\caption{Fit residuals for models with increasing mode content, for a scalar field ringdown (black, solid) with $\hat{n}=1 $ and $ A_{\hat{n}}=0.1 $ initial data. The minimal model providing a good fit to the data contains two fundamental modes and the first overtone (green). Fit residuals do not decrease significantly when higher overtones $ n>1 $ are added (pink), which we interpret as weak evidence for their presence in the data. The numerical noise is shown for reference (black, dashed). 
	} \label{fig:residuals}
\end{figure}

As an example, we provide the results of the fit for a perturbation with $\hat{n}=1$ and $A_{\hat{n}}=0.1$ in Table~\ref{tb:table_exc_Schw}. We also provide the fit mismatches in Table~\ref{tb:table_exc_Schw}, and plot the fit residuals as we add more modes to the model in Fig.~\ref{fig:residuals}. In all these diagnostic tools,
the evidence for overtones above the perturbing mode, $n>1 $ and including $ n=-1$, is very weak: the residuals do not improve significantly when adding these modes, and while the amplitudes of the first three modes remain consistent with the minimal model as overtones are added, the amplitudes of the higher tones are unstable: the $ n=2 $ mode amplitude doubles when adding the third overtone. Likewise, the mismatch decreases significantly when reaching the minimal model, but fluctuates around the minimum value as overtones are added. 

\setlength\extrarowheight{2pt}
\begin{table}[h]
\begin{ruledtabular}
	\begin{tabular}{ccccccc}
model &  $|\mathcal{M}|$ &  $A_{+0}$ &  $A_{-0}$ &  $A_1$ &  $A_2$ &  $A_3$ \\
	\colrule 
$+0$         &           0.93 &    3.1 $10^{-3}$ &    -- & -- & -- & -- \\ 
$\pm 0$      &           0.97  &    3.2 $10^{-3}$ &    6 $10^{-4} $& -- & -- & -- \\
$\pm 0, 1$ &           0.0040  &    3.3 $10^{-3}$ &    6 $10^{-4} $& 0.51 & -- & -- \\
$\pm 0, 1, 2$ &           0.0074  &    3.3 $10^{-3}$ &    6 $10^{-4}$ & 0.52 & 0.01 & -- \\ 
$\pm 0, 1,2,3$ &           0.0065  &    3.3 $10^{-3}$ &    6 $10^{-4}$ & 0.52 & 0.02 & 0.02  \\ 
\end{tabular}
\end{ruledtabular}
\caption{Mismatch between the data and the fit
	and the amplitudes of the overtones for the fits in Fig.~\ref{fig:residuals}.}\label{tb:table1}
\end{table}

\section{Higher overtone perturbations}\label{app:n2}

\begin{figure}[t]
	\centering
	\includegraphics[trim={.cm 0.6cm 2.2cm 2.1cm},clip,width=0.96\linewidth]{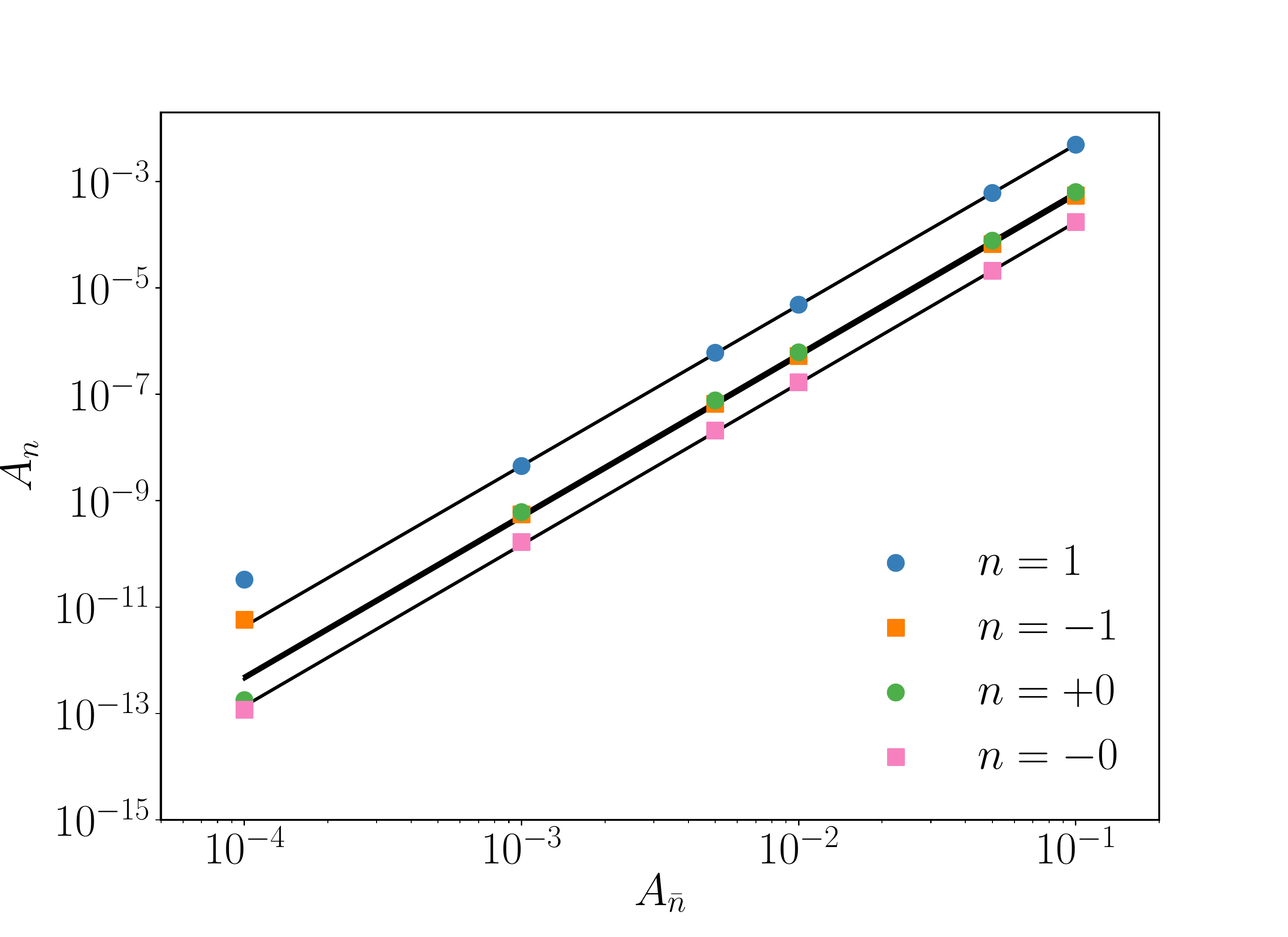} 
	\caption{Amplitude of excited modes in SAdS as a function of the $ \hat{n}=2 $ scalar perturbation amplitude, and a power-law fit (black). The fit returns exponents between $3.02$ and $3.03$.
	} \label{fig:amplitudes2}
\end{figure}

We confirmed that the amplitude of scalar nonlinear excitations in SAdS are cubic in the perturbation amplitude, for perturbation of any overtone number. Figure~\ref{fig:amplitudes2} shows the case of $ \hat{n}=2 $ perturbations. At higher $ \hat{n} $, the number of parameters to fit increases. Although the analysis becomes more challenging, the cubic relation is still distinguishable at high perturbation amplitudes.\\

\section{Angular momentum flux}\label{app:dJdM}
We compute the flux of angular momentum of a complex scalar field in a spherically symmetric black hole spacetime along the same lines of the energy flux calculation in Section~\ref{sec:theorybeyond}, see also Ref.~\cite{Dolan:2007mj}. The Killing vector associated with the $\varphi$ symmetry is given by $ \varphi^a = \partial_{\varphi} $ and defines a conserved angular momentum current, 
\begin{equation}\label{key}
J^a_{\rm ang} = T^{a b} \varphi_a \, .
\end{equation}
The integrated angular momentum flux across the event horizon reads
\begin{equation}\label{key}
\frac{d J}{d v} = \int_{\rm horizon} d\Omega \, n_a J^a_{\rm ang}  \, .
\end{equation}
Substituting the asymptotic form of the scalar field at the horizon, Eq.~\eqref{eq:scalar_horizon}, we find
\begin{equation}\label{key}
\frac{d J}{d v} = \frac{m \, \omega_{\rm R} }{8 \pi}  |A_{r_+}|^2 e^{2 v \omega_{\rm I}}\,
\end{equation}
Putting together this result and the energy flux computed in Eq.~\eqref{eq:fluxgeneric}, we obtain the generalization of the standard angular momentum-energy relation to perturbations with complex frequencies,
\begin{equation}\label{key}
d J = \frac{m \omega_{\rm R}}{|\omega|^2} dM \, . 
\end{equation}
Under the assumption of real frequency, the relation reduces to the well known $ d J = \frac{m}{\omega} dM  $.
%
\bibliography{Ref}

\end{document}